\documentclass[11pt]{article}

\usepackage[preprint]{acl}

\usepackage{times}
\usepackage{latexsym}

\usepackage[T1]{fontenc}

\usepackage[utf8]{inputenc}

\usepackage{microtype}

\usepackage{inconsolata}

\usepackage{graphicx}

\usepackage{listings}

\newcommand{\our}{\textsc{InteracSPARQL}}
\newcommand{\qaldt}{\texttt{QALD-10}}
\newcommand{\qald}{\texttt{QALD}}
\newcommand{\qaldn}{\texttt{QALD-9}}
\newcommand{\qaldnw}{\texttt{QALD-9-Wikidata}}
\newcommand{\qaldnd}{\texttt{QALD-9-DBpedia}}

\newcommand{\gpto}{\texttt{GPT-4o}}
\newcommand{\gptom}{\texttt{GPT-4o-mini}}
\newcommand{\cld}{\texttt{Claude-3.5-Sonnet}}
\newcommand{\qwt}{\texttt{Qwen-2.5-32B}}
\newcommand{\qwf}{\texttt{Qwen-2.5-14B}}

\usepackage{float} 
\usepackage{enumitem}
\usepackage{xcolor}
\usepackage{mdframed}
\usepackage{amsmath}
\usepackage{tcolorbox}
\usepackage{ragged2e}
\usepackage{caption}

\usepackage{booktabs}
\usepackage{subcaption}
\usepackage{siunitx} 
\usepackage{multirow}
\usepackage{geometry}
\usepackage{circledsteps}
\usepackage[linesnumbered,ruled,vlined]{algorithm2e} 
\usepackage{amsmath} 
\usepackage{amsthm}
\newtheorem{definition}{Definition}
\definecolor{queryblue}{RGB}{0,102,204}      
\definecolor{bracketblue}{RGB}{0,153,153}    
\definecolor{patternpurple}{RGB}{204,0,204}  
\definecolor{bracketpurple}{RGB}{255,102,0}  
\definecolor{explanationgray}{RGB}{120,120,120} 

\definecolor{bggray}{HTML}{F9FAFB}
\definecolor{textgray}{HTML}{4B4B4B} 
\definecolor{darkgray}{HTML}{2F4F4F}

\definecolor{vargreen}{RGB}{34, 139, 34}
\definecolor{prefixbrown}{RGB}{139, 69, 19}
\definecolor{bracketgreen}{RGB}{0, 153, 51}    
\definecolor{lightgray}{RGB}{248,248,248}

\lstdefinelanguage{JSON}{
    basicstyle=\ttfamily\footnotesize,
    showstringspaces=false,
    breaklines=true,
    frame=single,
    backgroundcolor=\color{lightgray},
    literate=
     *{0}{{{\color{blue}0}}}{1}
      {1}{{{\color{blue}1}}}{1}
      {2}{{{\color{blue}2}}}{1}
      {3}{{{\color{blue}3}}}{1}
      {4}{{{\color{blue}4}}}{1}
      {5}{{{\color{blue}5}}}{1}
      {6}{{{\color{blue}6}}}{1}
      {7}{{{\color{blue}7}}}{1}
      {8}{{{\color{blue}8}}}{1}
      {9}{{{\color{blue}9}}}{1}
      {:}{{{\color{black}{:}}}}{1}
      {,}{{{\color{black}{,}}}}{1}
      {\{}{{{\color{black}{\{}}}}{1}
      {\}}{{{\color{black}{\}}}}}{1}
      {[}{{{\color{black}{[}}}}{1}
      {]}{{{\color{black}{]}}}}{1},
    escapeinside={(*@}{@*)},
    morecomment=[l]{//},
    commentstyle=\color{gray}\ttfamily,
}

\tcbset{
  mysymbox/.style={
    width=\linewidth,            
    sharp corners,
    boxrule=.8pt,                
    colframe=blue!75!black,
    colback=white,
    coltitle=white,
    fonttitle=\bfseries,
    left=5pt,right=5pt,top=5pt,bottom=5pt, 
  }
}

\lstdefinestyle{sparql-vivid}{
  basicstyle=\scriptsize\ttfamily,
  escapeinside={(*@}{@*)},
  breaklines=true,
  breakatwhitespace=true,
  columns=fullflexible,
  frame=single,
  backgroundcolor=\color{gray!5},
  xleftmargin=0pt,
  xrightmargin=0pt,
  framexleftmargin=3pt,
  framexrightmargin=3pt,
  aboveskip=8pt,
  belowskip=8pt,
  postbreak=\mbox{\textcolor{explanationgray}{$\hookrightarrow$}\space},
}
\lstdefinelanguage{nle}{
    basicstyle=\ttfamily\footnotesize,
    numbers=left,
    numberstyle=\tiny\color{gray},
    stepnumber=1,
    numbersep=5pt,
    showstringspaces=false,
    breaklines=true,
    frame=single,
    backgroundcolor=\color{gray!5},
    literate=
     *{\{}{{{\textbf{\textcolor{black}{\{}}}}}{1}
      {\}}{{{\textbf{\textcolor{black}{\}}}}}}{1}
      {[}{{{\textbf{\textcolor{bracketgreen}{[}}}}}{1}
      {]}{{{\textbf{\textcolor{bracketgreen}{]}}}}}{1}
      {<}{{{\textbf{\textcolor{bracketblue}{<}}}}}{1}
      {>}{{{\textbf{\textcolor{bracketblue}{>}}}}}{1}
      {(}{{{\textbf{\textcolor{bracketpurple}{(}}}}}{1}
      {!)}{{{\textbf{\textcolor{bracketpurple}{)}}}}}{1}
      {:}{{{\textbf{\textcolor{black}{:}}}}}{1}
      {,}{{{\textbf{\textcolor{black}{,}}}}}{1}
      {"Overall_NL_explanation"}{\textbf{\textcolor{queryblue}{"Overall NL explanation"}}}{23}
      {"Query_Type"}{\textbf{\textcolor{queryblue}{"Query Type"}}}{11}
      {"Variables"}{\textbf{\textcolor{queryblue}{"Variables"}}}{10}
      {"Patterns"}{\textbf{\textcolor{queryblue}{"Patterns"}}}{10}
      {"Prefixes"}{\textbf{\textcolor{queryblue}{"Prefixes"}}}{10}
      {"Description"}{\textbf{\textcolor{vargreen}{"Description"}}}{12}
      {"Variable"}{\textbf{\textcolor{vargreen}{"Variable"}}}{9}
      {"Module"}{\textbf{\textcolor{patternpurple}{"Module"}}}{7}
      {"Prefix"}{\textbf{\textcolor{prefixbrown}{"Prefix"}}}{7}
      {"SPARQL"}{\textbf{\textcolor{black}{"SPARQL"}}}{7}
      {"Explanation"}{\textbf{\textcolor{explanationgray}{"Explanation"}}}{11},
}
\title{\our: An Interactive System for SPARQL Query Refinement Using Natural Language Explanations}

\author{Xiangru Jian \\
 University of Waterloo \\
  \texttt{xiangru.jian@uwaterloo.ca} \\\And
  Zhengyuan Dong \\
 University of Waterloo \\
  \texttt{zhengyuan.dong@uwaterloo.ca} \\\AND
  M. Tamer Özsu \\
   University of Waterloo \\
   \texttt{tamer.ozsu@uwaterloo.ca}
  }

\begin{document}

\maketitle
\begin{abstract}

In recent years, querying semantic web data using SPARQL has remained challenging, especially for non-expert users, due to the language’s complex syntax and the prerequisite of understanding intricate data structures. To address these challenges, we propose \our, an interactive SPARQL query generation and refinement system that leverages natural language explanations (NLEs) to enhance user comprehension and facilitate iterative query refinement. \our\ integrates LLMs with a rule-based approach to first produce structured explanations directly from SPARQL abstract syntax trees (ASTs), followed by LLM-based linguistic refinements. Users can interactively refine queries through direct feedback or LLM-driven self-refinement, enabling the correction of ambiguous or incorrect query components in real time. We evaluate \our\ on standard benchmarks (\qaldn\ and \qaldt), demonstrating significant improvements in query accuracy, explanation clarity, and overall user satisfaction compared to baseline approaches. Our experiments further highlight the effectiveness of combining rule-based methods with LLM-driven refinements to create more accessible and robust SPARQL interfaces.

\end{abstract}
\section{Introduction}

Querying RDF (Resource Description Framework) data with SPARQL has long been challenging for users lacking substantial technical expertise~\cite{li-etal-2023-shot,icde21_9458607,Arenas:2017aa,vldb16_Diaz:2016,Mohamed:2022vu}. Although SPARQL is a powerful language for working with linked data, its complex syntax, coupled with the need to understand underlying data organization without a clear schema, often creates a steep learning curve. Moreover, RDF resources are typically identified by \emph{Internationalized Resource Identifiers (IRIs)}, a superset of URIs that accommodates a broader range of characters, which can further complicate query formulation for less experienced users. The difficulty of using SPARQL has been identified as a major issue in user surveys \cite{www19_Bonifati:2019aa,vldb18_BonifatiMT17,ariasusewod2011}. As organizations increasingly adopt RDF-based systems in domains such as life sciences (e.g., Bio2RDF~\cite{bio2rdf}), social graphs, and knowledge bases (e.g., Wikidata~\cite{wikidata} and DBpedia~\cite{morsey:2011vn}), the need for user-friendly interfaces grows~\cite{vldb21_Helal:2021}.

\begin{figure}
    \centering
    \includegraphics[width=\linewidth]{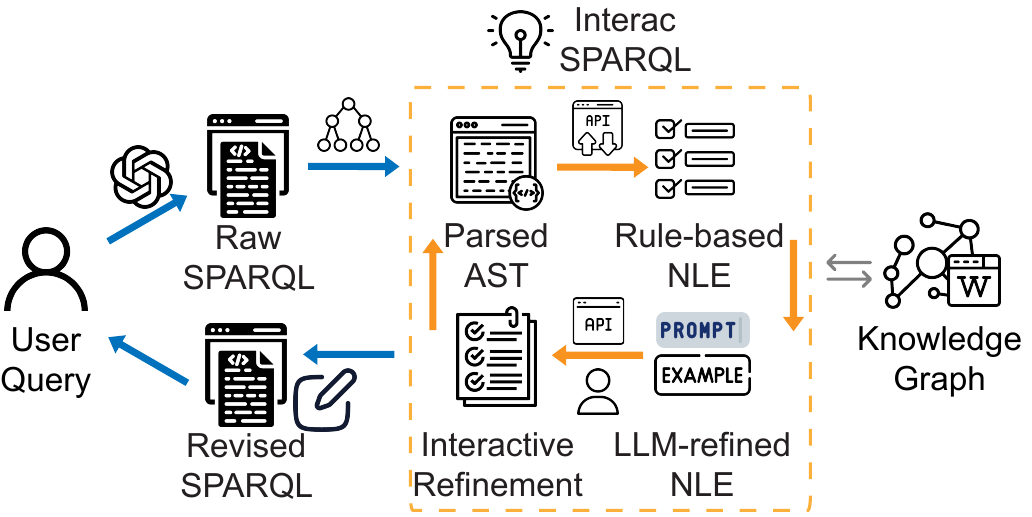}
    \caption{The overview of \our\ .}
    \label{fig:overview}
\end{figure}

While similar challenges also apply to other query languages such as SQL, our focus on SPARQL is motivated by several distinct factors. SPARQL's extensive use in querying RDF-based linked datasets differs from SQL, particularly due to its graph-oriented query structure, extensive reliance on IRIs, and semantic web applications. Moreover, SPARQL’s complexity, which includes advanced constructs like property paths and complex navigational constructs, makes it particularly challenging for non-experts, as indicated in an overview of SPARQL evaluation~\cite{cohen-kim-2013-evaluation}.

Given these challenges, natural language to SPARQL query (NL2SPARQL) provides the most natural, easy, and practical interface for users to write SPARQL queries. A typical NL2SPARQL pipeline involves interpreting a user’s natural language question, identifying and grounding entities and properties in the target knowledge graph, and composing a complete SPARQL query that reflects the user’s intent. However, existing NL2SPARQL systems suffer from relatively low accuracy (often only 20–40\% F1 on challenging benchmarks)~\cite{li-etal-2023-shot,jiang2023structgptgeneralframeworklarge,liu2024spinachsparqlbasedinformationnavigation,10662970,Angles:2022aa}, and provide little transparency or explainability, making it difficult for users to verify or refine the generated queries. Bridging natural language to SPARQL is particularly challenging because SPARQL is a low-resource language. The performance of these systems is likely to improve with human-in-the-loop assistance based on meaningful explanation and feedback from the system.




To address these gaps, recent NL2SPARQL research has explored \textit{one-turn SPARQL generation}~\cite{sigmod23_Omar:2023aa,jiang2023structgptgeneralframeworklarge,xie-etal-2022-unifiedskg,yu2023decaf}, which directly transforms a natural language question into a complete SPARQL query in a single step without interactive refinement. Despite its convenience, most works in this approach typically yields low accuracy due to the complexity and ambiguity of user queries, and offers no means for users to inspect or correct the query logic. For some specialized models~\cite{SGPT,xie-etal-2022-unifiedskg} can perform well on certain benchmark, they all need intensive training and thus lack of generalization ability to others.

Building on these efforts, a number of tools have adopted an \textit{interactive SPARQL query refinement} paradigm~\cite{icde21_9458607,icde18_8509280,OCHIENG2020100024,vldb12_LetelierPPS12}. Such tools let users iteratively adjust their queries—often by inspecting intermediate results or invoking domain‐specific heuristics—to converge on the desired logic. However, they typically support only a subset of SPARQL’s full feature set (for example, omitting property paths or complex filtering) and rarely supply clear, structured natural‐language explanations of each query component. As a result, novice users still struggle with opaque syntax, and even expert users find the limited feature coverage and feedback mechanisms leave critical queries unrefined.

In this paper, we introduce \textbf{\our}, an interactive SPARQL query generation and refinement system that addresses these challenges with a holistic design (Fig.~\ref{fig:overview}). \our\ is intended to be used either as a companion to NL2SPARQL models to assist in generating correct queries, or as part of training systems that help users learn SPARQL on all RDF-based systems with minimal adaptation needed. Our key contribution is to develop an interactive tool that assists users in refining their queries by providing \textbf{natural language explanations (NLE)} for each section of a SPARQL query.

By combining a rule-based method for deriving a structured, deterministic explanation from the original SPARQL query (or its abstract
syntax tree, AST) with a subsequent LLM-based refinement step, \our\ yields explanations that are both accurate and linguistically polished. The interactive refinement is done either by direct oversight or through an LLM ``self-refinement''. As a result, corrections to mislabeled entities, ambiguous filters, or incomplete clauses are achieved in real time. Furthermore, \our\ supports users who wish to author SPARQL directly by guiding them step-by-step through query construction, providing continuous, targeted feedback and effectively serving as an interactive learning guide to SPARQL. This educational dimension is woven throughout the paper: our methodology formalizes a step-by-step construction mode, our experiments evaluate not only accuracy but also the usefulness of explanations as a learning aid, and our human study explicitly measures the extent to which users gain confidence and understanding of SPARQL syntax through interactive guidance.

The design of \our\ tackles several core difficulties. One is how to systematically split SPARQL syntax into interpretable modules—such as \texttt{Basic Graph Patterns (BGPs)}, \texttt{FILTER} clauses, or advanced constructs like \texttt{GROUP BY}—so that users can see and modify each component. Another challenge is ensuring that iterative refinements stay aligned with the user’s goals, especially when advanced features (like subqueries or property paths) are involved. Additionally, controlling the interaction to minimize the number of required user revisions is also essential, since excessive revisions often deter users due to frustration and inefficiency. To address these challenges, our approach couples incremental improvements with dynamic \textit{tool-based entity and property lookups}, letting the user or an LLM swiftly resolve uncertain references and ambiguities in real time. Ultimately, the synergy of NLEs and multi-round refinement aims to make SPARQL more transparent and approachable for novices, while retaining robust capabilities for domain experts who need complex queries.

Our work presents a fresh perspective on integrating \textit{modular explanation} and \textit{iterative refinement} into the SPARQL generation pipeline. Specifically, we:
\begin{enumerate}[leftmargin=*]
     \item propose a \textbf{two-stage NLE framework}, which systematically parses SPARQL queries into structured explanations via rule-based AST analysis, followed by linguistic refinement using LLMs. This design ensures accuracy, interpretability, and fluency of generated explanations.


    \item develop an \textbf{interactive query construction and refinement framework} that (i) primarily supports iterative improvements via direct user input or automated LLM-driven self-refinement, aligning queries closely with user intent; (ii) also guides users through each step of SPARQL authoring, providing continuous, targeted feedback as an educational aid. (iii) and can be easily adopted by any RDF-based system without any training.

    \item introduce a \textbf{dynamic, tool-assisted entity and property linking mechanism}, invoked during interactive refinement, which efficiently resolves domain-specific ambiguities and contributes to stable convergence toward correct and precise IRIs.

    \item conduct \textbf{comprehensive experimental evaluations} using the \qald\ benchmarks, demonstrating significant improvements in query accuracy, iterative refinement capability, and overall user satisfaction compared to state-of-the-art methods. Importantly, we further validate the quality and practical utility of our explanations through a human evaluation study, confirming substantial improvements in clarity, completeness, correctness, and utility.

\end{enumerate}

Additional technical details and extended examples are provided in a supplementary Appendix (denote as Appendix in the paper) available at: https://bit.ly/interacSparql\_app.

\section{Background} \label{sec:background}


RDF and SPARQL are foundational technologies of the Semantic Web, a framework that aims to create a more intelligent, interconnected web. RDF, introduced by the World Wide Web Consortium (W3C), is a standard model for data representation and interchange on the web, facilitating the integration and sharing of information across different domains~\cite{Ali:2022vx,Wylot:2018aa,vldb17_AbdelazizHKK17,angles:2008aa,pods11_arenasp11,icde15_7113413,icde21_9458632}. RDF represents data as triples, consisting of a subject, predicate, and object (usually represented as $(s,p,o)$), which together form a graph structure. This simple yet flexible model allows for the expression of complex data relationships and supports interoperability between heterogeneous data sources. Formally, an RDF dataset and an RDF graph can be defined as in Definitions \ref{def:rdf} and \ref{def:rdfgraph}.

\begin{definition}\label{def:rdf}
Let $\mathcal{I}, \mathcal{B}, \mathcal{L},$ and $\mathcal{V}$ denote the sets of all URIs, blank nodes, literals, and variables, respectively. A triple $(s,p,o) \in (\mathcal{I} \cup \mathcal{B}) \times \mathcal{I} \times (\mathcal{I} \cup \mathcal{B} \cup \mathcal{L})$ is an \textit{RDF triple}, where $s$, $p$ and $o$ are called subject, property (or predicate) and object. A set of RDF triples form a \textit{RDF dataset}.
\end{definition}

\begin{definition}\label{def:rdfgraph}
A \emph{RDF graph} $G=\langle V,L_V, f_{V}, E,L_E, f_{E} \rangle$ is a six-tuple , where

\begin{enumerate}[leftmargin=*]
	\item $V=V_r\cup V_l$ is a collection of vertices that correspond to all subjects and objects in RDF data, where $V_r$ and $V_l$ are collections of resource vertices and literal vertices, respectively.
	\item $L_V$ is a collection of vertex labels. 
	\item A \emph{vertex labeling function} $f_{V}: V \rightarrow L_{V}$ is a bijective function that assigns to each vertex a label. The label of a vertex $u \in V_l$ is its literal value, and the label of a vertex $u \in V_r$ is its corresponding URI or the blank node identifier.
	\item $E=\{\overrightarrow{u_1,u_2}\}$ is a collection of directed edges that connect the corresponding subjects and objects.
	\item $L_E$ is a collection of edge labels. 
	\item An \emph{edge labeling function} $f_{E}: E \rightarrow L_{E}$ is a bijective function that assigns to each edge a label. The label of an edge  $e \in E$ is its corresponding predicate (or called property).
	
\end{enumerate}
An edge $\overrightarrow{u_1,u_2}$ is an \emph{attribute property}  edge if $u_2 \in V_l$; otherwise, it is a \emph{link} edge.
\end{definition}

SPARQL, also standardized by the W3C\cite{w3c:2006aa,angles:2008aa,perez2009,pods11_arenasp11,hartig2009a,harris:aa,esws12_hartig12}, is a powerful query language designed to retrieve and manipulate RDF data. An intuitive definition of SPARQL is given in Definition \ref{def:sparql}.

\begin{definition}\label{def:sparql}
A SPARQL query typically includes five parts: 
\begin{enumerate}[leftmargin=*]
    \item The \textit{prefix declarations} are used to simplify and shorten the URIs (Uniform Resource Identifiers) that are commonly used in RDF data.  Prefix declarations allow the user to define a shorthand notation (a prefix) for a namespace URI, making the query more readable and easier to write.
    \item The \textit{output part} of a SPARQL can be in the form of a table of values of variables (\texttt{SELECT}), or in a RDF graph specified by a graph template substituting for the variables by each query solution in the graph template (\texttt{CONSTRUCT}), or testing whether or not a query pattern has a solution (\texttt{ASK}). 
    \item The \textit{dataset definition} refers to the specification of the RDF dataset that a query operates on and is identified in the \texttt{FROM} clause. 
    \item The \textit{graph pattern matching part} is specified in the \texttt{WHERE} clause and includes a set of triple patterns to be matched as well as \texttt{OPTIONAL}, \texttt{UNION} and \texttt{FILTER} operators. The data source to be matched is specified by \texttt{FROM} in this part. 
    \item The \textit{solution modifier part} includes projection, distinct, order and limit operators defined over the graph pattern matching results. 
 
\end{enumerate}
\end{definition}

If the graph pattern matching part consist of only triple patterns (no \texttt{OPTIONAL}, \texttt{UNION} or \texttt{FILTER}), this is called \textit{basic graph pattern} (BGP) query.

SPARQL has undergone significant enhancements since its original release in 2008. The current version, SPARQL 1.1 \cite{sparql11-overview},  includes support for updates, property paths, aggregates, subqueries, negation, and nested queries, among other features. 


\captionsetup[lstlisting]{name=Example} 

\begin{lstlisting}[basicstyle=\scriptsize\ttfamily,caption={A SPARQL Query Example in \qaldt}, label={lst:query_exp}, float]
SELECT ?tvShow WHERE {
  ?tvShow wdt:P31 wd:Q5398426;  
  ?tvShow wdt:P161 wd:Q23760;   
  ?tvShow wdt:P2437 ?seasons;   
  ?tvShow wdt:P580 ?startDate.  
  FILTER(?seasons = 4)
  FILTER(YEAR(?startDate) = 1983)}
\end{lstlisting}

The semantics of SPARQL are based on graph pattern matching using homomorphism, where the query engine searches for graph patterns specified in the \texttt{WHERE} clause against the RDF data. If the pattern matches, the variables in the query are bound to corresponding values from the RDF graph, and the query result is constructed accordingly. SPARQL supports various forms of graph pattern matching, including BDPs, optional patterns, and union patterns.


To illustrate the syntax and semantics of SPARQL, consider the query in Example~\ref{lst:query_exp}. This query retrieves the answer to the natural language question ``\textit{What is the TV-show that starred Rowan Atkinson, had 4 seasons and started in 1983?}''. This is formulated in SPARQL query where the answer is bound to the variable \texttt{?tvShow} that satisfies four conditions: they must be \textbf{an instance of} (\texttt{wdt:P31}) the \textbf{television series} (\texttt{wd:Q5398426}) via the triple pattern \texttt{?tvShow wdt:P31 wd:Q5398426}; they must feature \textbf{Rowan Atkinson} (\texttt{wd:Q23760}) as a \textbf{cast member} (\texttt{wdt:P161}) via \texttt{?tvShow wdt:P161 wd:Q23760}; they must have their \textbf{number of seasons} (\texttt{wdt:P2437}) bound to \texttt{?seasons} via \texttt{?tvShow wdt:P2437 ?seasons}; and they must have their \textbf{start date} (\texttt{wdt:P580}) bound to \texttt{?startDate} via \texttt{?tvShow wdt:P580 ?startDate}. The first \texttt{FILTER} clause \texttt{FILTER(?seasons = 4)} ensures only shows with exactly four seasons are considered, while the second \texttt{FILTER} clause \texttt{FILTER(YEAR(?startDate) = 1983)} restricts results to those that began in the year 1983. Finally, the \texttt{SELECT} clause returns each matching \texttt{?tvShow}.


\section{\our\ System} \label{sec:method}

\begin{figure*}
    \centering\includegraphics[width=1\textwidth]{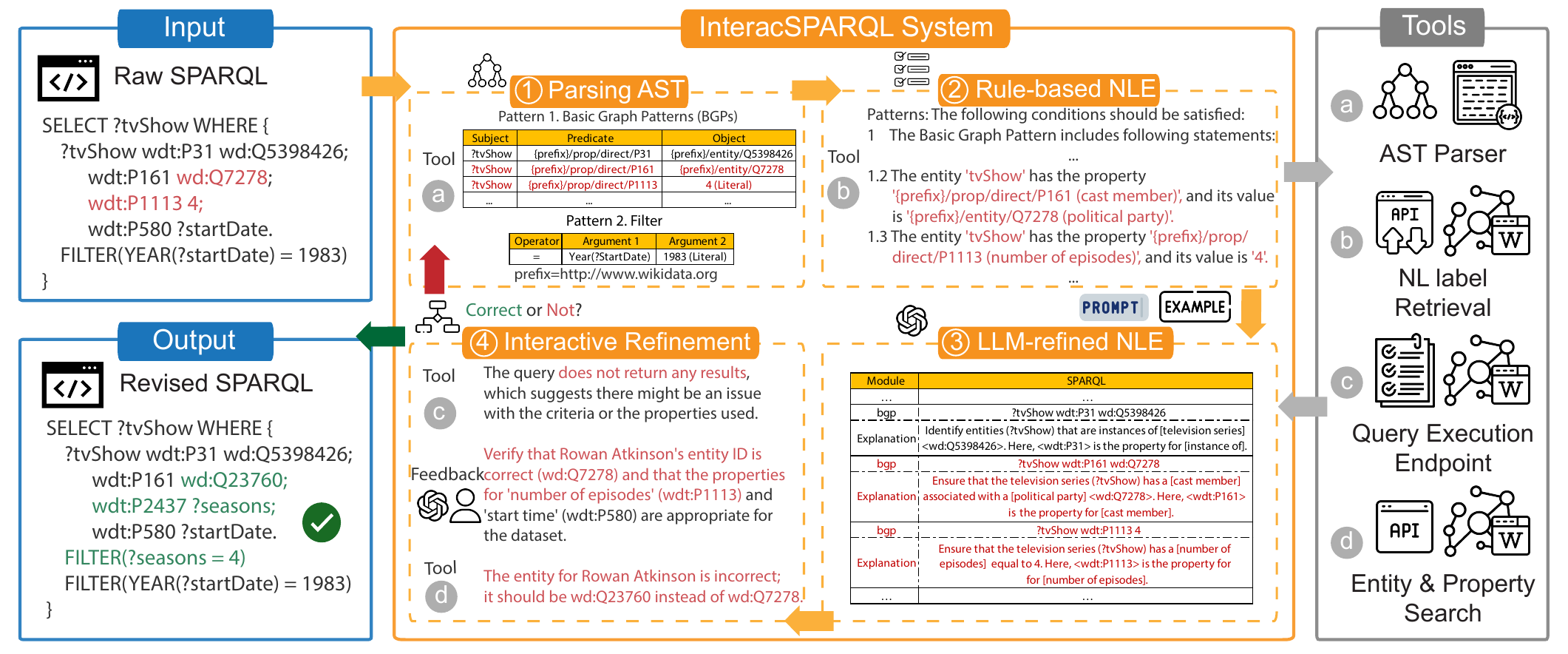}
    \caption{The proposed pipeline for \our. The input is the raw generation of GPT-4o over the natural language question: \textit{What is the TV-show that starred Rowan Atkinson, had 4 seasons and started in 1983?}, which is incorrect. The output query is produced by \our\ (Example~\ref{lst:query_exp}) and is identical to the ground truth.}
    \label{fig:pipeline}
\end{figure*}

In this section, we describe \our\ that implements a pipeline to generate SPARQL query explanations and perform iterative query refinement upon them. \our\ addresses critical shortcomings of existing systems, namely, the lack of transparent explanations and a robust mechanism for iterative query improvement. Further comparisons with existing systems are discussed in Section~\ref{sec:related}.

\our\ pipeline is depicted in Figure~\ref{fig:pipeline}. The steps involved in this process are as follows: \Circled{1} parsing the original SPARQL into a structured JSON-based Abstract Syntax Tree (AST), thus providing a machine-readable blueprint for the query; \Circled{2} apply the AST to produce concise, hierarchical rule-based NLEs with IRIs replaced by human-readable labels via on-demand lookups; \Circled{3} Use an LLM with specially curated few-shot examples to refine rule-based NLEs into a fluent, structured JSON explanation detailing overall intent, query type, variable descriptions, clause modules, and prefix clarifications; \Circled{4} Use an interactive query refinement loop to flag problematic clauses or stale entities based on LLM-refined NLEs, gather targeted feedback (from a human or via self-refinement), apply incremental fixes through RDF interaction tools, and repeat until the query’s results and explanation are semantically accurate and align with the user’s intent.

In the following, we focus on steps \Circled{2}, \Circled{3}, and \Circled{4}, since these are the technically interesting ones. Step \Circled{1} is essential but technically well understood. Furthermore, rather than discuss each of these steps one by one, we take a more integrated approach and discuss the two challenges that these steps address: Natural Language Explanation (Section \ref{subsec:nle}) and Interactive Query Refinement (Section \ref{subsec:interactive-refinement}). Within this organization, Section \ref{subsec:rule-based-nle} explains \Circled{2}, Section \ref{subsec:llm-nle} focuses on \Circled{3}, and Section \ref{subsec:interactive-refinement} covers \Circled{4}. Other details (e.g. details for step \Circled{1}) can be found in Section in Appendix.

\subsection{Natural Language Explanation}
\label{subsec:nle}

Producing clear, accurate, and intuitive explanations for SPARQL queries poses significant challenges, particularly when the objective is to support iterative refinement by both human users and language models. To address this, we introduce a two-stage approach (illustrated in Fig. ~\ref{fig:pipeline}) that carefully balances clarity, accuracy, and computational efficiency. 

Initially, we employ a structured, rule-based technique to extract precise, deterministic explanations directly from AST (Section~\ref{subsec:rule-based-nle}). This foundational step ensures each SPARQL query element is clearly represented, thereby providing an interpretable, reliable baseline. Subsequently, these structured explanations are passed to an LLM, enriched with carefully selected few-shot examples (Section~\ref{subsec:llm-nle}). Leveraging the rule-based foundation allows the LLM to concentrate on linguistic refinement—enhancing readability and capturing nuanced contextual insights—without sacrificing factual accuracy.

\subsubsection{Extracting Rule-Based NLE from AST}
\label{subsec:rule-based-nle}

Our method systematically traverses AST's hierarchical structure and transforms each AST node into succinct, human-readable statements clearly articulating its purpose. For example, in \textbf{step \Circled{2}} of Fig.~\ref{fig:pipeline}, the complex query components like Basic Graph Patterns (BGPs) and \texttt{FILTER} clauses are decomposed into straightforward sentences (e.g., ``The entity ``tvShow'' has the property \texttt{wdt:P1113} (number of episodes), and its value is ``4''.''). This ensures even complex SPARQL patterns remain transparent and understandable. The details of this process is demonstrated by Example 3 in the Appendix, which is the rule-based NLE of Example~\ref{lst:query_exp}. 

\paragraph{Leveraging Hierarchical Structure for Clarity.}Rather than flattening the query into linear explanations, we maintain AST's hierarchical form. Each node and sub-node explanation explicitly highlights its semantic relationship within the query structure. This design helps users identify precisely where modifications may be required, significantly facilitating targeted query refinement.

\paragraph{Contextual Enrichment of Identifiers by Label Search.}
Recognizing the challenge posed by opaque IRIs, we enhance explanations by integrating labels from underlying knowledge bases (such as Wikidata). Transforming identifiers like \texttt{wd:Q5398426} into descriptive names like \textit{``television series''} significantly reduces cognitive load, enabling users to recognize entities and predicates immediately.


This structured, rule-based explanation acts as a stable intermediate form, ensuring consistency and reducing the risk of LLM hallucinations. By providing a clear semantic backbone, the LLM can focus purely on enhancing fluency, nuance, and readability—key aspects that directly improve interpretability and user comprehension.

\subsubsection{Refinement of NLE via LLMs: Structured Guidance and Enriched Narrative}
\label{subsec:llm-nle}

Building upon the concise rule-based explanations, we leverage LLMs to generate linguistically polished and contextually enriched narratives. This is \textbf{step \Circled{3}} in Fig.~\ref{fig:pipeline}. This refinement process critically integrates two essential components: (1) the structured semantic and hierarchical information extracted from the AST and the rule-based NLE, ensuring accurate preservation of query structure and content; and (2) carefully designed few-shot examples presented in an accessible, transparent, and hierarchical format, serving as good references to guide the LLM towards coherent, structured NLEs.

\paragraph{Structured Input for Controlled Refinement Workflow.}
The LLM refinement stage explicitly leverages structured information from the AST-derived, rule-based explanations, combined with rigorously selected few-shot examples. Rather than interpreting the original SPARQL query anew, the model's role is constrained to refining validated, structured content. This explicit guidance minimizes semantic inaccuracies, reduces the likelihood of hallucinations, and facilitates domain-specific insights, thereby maintaining precise alignment with the original query intent.

\paragraph{Customized JSON-based Output Format.}
To reinforce clarity and maintain coherence across refinement iterations, we adopt a structured, JSON-based output format that systematically reflects essential query elements. The resulting explanations encompass distinct modular sections: \textbf{(1) Overall NL explanation}, summarizes the primary objective and intent of the query, providing immediate context; \textbf{(2) Query type}, explicitly identifies the query format (e.g., \texttt{SELECT}, \texttt{ASK}), clarifying its functional purpose within the dataset; \textbf{(3) Variables}, clearly articulate each query variable's role, connecting it explicitly to broader retrieval logic; \textbf{(4) Modules (graph patterns and clauses)}, present significant query components alongside corresponding SPARQL code snippets and explanatory narratives, preserving the hierarchical query structure; \textbf{(5) Advanced clauses}, highlight advanced query features (e.g., \texttt{GROUP BY}, \texttt{ORDER BY}), clarifying their effects on query results; \textbf{(6) Prefixes}, provide optional contextual explanations of prefixes to clarify namespace conventions and minimize confusion.

By the end of step \Circled{3}, each terse, AST-derived clause is transformed into a module in a JSON-formatted explanation, pairing each SPARQL fragment with a polished natural-language description and organizing content into explicit sections for variables, graph patterns, filters, and prefixes. An excerpt of this partial NLE is shown in step \Circled{3} of Fig.~2. For example, the pattern \texttt{{?tvShow wdt:P1113 4}}
is rendered as “Ensure the television series has exactly four episodes,” replacing the raw IRI with its human-readable label and employing concise, engaging phrasing. The full explanation appears as Example 4 in the Appendix. These structured modules markedly improve clarity and usability, enabling efficient iterative refinement for both novices and experts.


\subsection{Interactive Query Refinement}
\label{subsec:interactive-refinement}

Extending the foundation established by the NLE framework (Section~\ref{subsec:llm-nle}), we employ an \textit{interactive query refinement} process that aligns SPARQL queries with the user’s original question. This is \textbf{step \Circled{4}} in Fig.~\ref{fig:pipeline}. While the system can readily incorporate direct user feedback, we also offer a \textit{self-refinement} mode in which an LLM simulates user suggestions for automated evaluation. In this section, we outline the key steps of the refinement loop, explain how the NLE underpins each iteration, and highlight a tool-based entity/property search mechanism that reduces the domain knowledge burden for query authors. Our current prototype implementation features these tools for popular knowledge graphs like Wikidata and DBpedia, but the same methodology can be adapted to other semantic datasets with minimal modification.

\subsubsection{Motivation for Tool-based Entity and Property Search}
When writing or refining SPARQL queries, it is often necessary to reference exact entity and property URIs
Users or LLMs may not recall these IRIs offhand, leading to guesswork and errors. To address this challenge, we incorporate dedicated search tools that the LLM can invoke on demand. These functions query the target knowledge graph’s API or index to identify proper IRIs for entities (e.g., ``Rowen Atkinson'') or properties (e.g., ``start date''). By delegating entity/property linking to a well-defined utility, the iterative refinement loop becomes more convenient and robust, relieving users and the LLM of low-level domain details.

\begin{algorithm}[t]
\small
\LinesNumbered 
\caption{Interactive Query Refinement Algorithm}
\label{alg:interactive-refine}
\DontPrintSemicolon
\SetKwInOut{Input}{Input}
\SetKwInOut{Output}{Output}
\SetKwFunction{execute}{execute}
\SetKwFunction{genNLE}{generateOrUpdateNLE}
\SetKwFunction{isConsistent}{isConsistent}
\SetKwFunction{getFeedback}{getFeedback}
\SetKwFunction{applyFeedback}{applyFeedback}
\SetKwFunction{toolCall}{toolCall}

\Input{
$U$: Natural Language Question (i.e. user's intent) \\
$\mathcal{Q}$: Initial SPARQL query \\
$K$: Target knowledge graph \\
$N$: Maximum refinement iterations
}
\Output{
$\mathcal{Q}^{\ast}$: Refined SPARQL query aligned with user intent
}

\vspace{2pt}
\BlankLine

$i \leftarrow 0$\;
\While{$i < N$}{
    \tcp{1. Explain \& Validate}
    $\mathcal{R} \leftarrow \execute(\mathcal{Q}, K)$\;
    $\mathrm{NLE} \leftarrow \genNLE(\mathcal{Q}, \mathcal{R})$\;
    \If{\isConsistent(Q, $\mathcal{R}$, NLE, U)}{
        \textbf{break} \tcp*{Stop if results align with user's question}
    }

    \tcp{2. Evaluate \& Provide Feedback}
    $\mathrm{fb} \leftarrow \getFeedback(\mathcal{Q}, \mathrm{NLE}, \mathcal{R})$\;

    \tcp{3. Refine the Query}
    \If{\text{feedback indicates incorrect entity or property}}{
        \toolCall(\text{search function for entity/property}) \tcp*{The LLM retrieves the appropriate IRI}
    }
    $\mathcal{Q} \leftarrow \applyFeedback(\mathcal{Q}, \mathrm{fb}, \mathrm{NLE})$\;

    $i \leftarrow i + 1$\;
}

\Return $\mathcal{Q}^{\ast} \leftarrow \mathcal{Q}$\;
\end{algorithm}

\subsubsection{Detailed Workflow}
Algorithm~\ref{alg:interactive-refine} illustrates the workflow of the \textit{interactive query refinement} through four primary stages. (also refer to Figure~\ref{fig:pipeline}): 


\begin{enumerate}[leftmargin=*]
    \item \textbf{Stage 1: Explain and Validate (Alg.~\ref{alg:interactive-refine}, lines 3–4).}  
    The system executes the SPARQL query on the chosen knowledge graph and collects results. Concurrently, it creates or updates the NLE to reflect the query’s logical structure, enumerating triple patterns, filters, and so on. So the output of this stage will be the execution results and NLE of the given query. Take the input query in Fig.~\ref{fig:pipeline} for example (denoted as \textbf{Raw Query} for convenience), the execution result is empty, the NLE is like the one \textbf{step \Circled{3}}).
    

    \item \textbf{Stage 2: Evaluate and Provide \textit{Feedback} (lines 5–7).}  
    Should the query’s output prove not to accurately reflect the user's intention (i.e. the natural language question given), a feedback mechanism (either a human user or an LLM) identifies possible reasons for the mismatch, such as a wrong property URI. This feedback delineates which segments of the query (variables, filters, patterns) demand revision. If the current outputs already meet the user’s intention, feedback will also be given to indicate no further effort is required. The feedback for \textbf{Raw Query} is \textit{Verify that Rowan Atkinson's entity ID is correct (wd:Q7278) and that the properties for 'number of episodes' (wdt:P1113) and 'start time' (wdt:P580) are appropriate for the dataset.}

    \item \textbf{Stage 3: Refine the Query (lines 8–10).}  
    Using the feedback, the system selectively updates the query. If the feedback indicates that a particular entity or property is missing or erroneous, the LLM may invoke the entity/property search tool to perform an on-demand search and discover the correct IRI. The feedback of \textbf{Raw Query} evokes two searches, one for the entity ``Rowan Atkinson'' and the other for properties ``number of episode'' and ``start time''. It turns out that the first two searches indicate errors in the \textbf{Raw Query}, giving IRI of ``\texttt{wd:23760}'' and ``\texttt{wdt:2437}'', respectively. The refined query modifies only the problematic references, preserving previously validated logic. Reflecting on \textbf{Raw Query}, IRI of entity ``Rowan Atkinson'' gets updated and the property ``number of season'' (\texttt{wdt:1113}) gets replaced by ``number of episode'' (\texttt{wdt:2437}). Also, an extra \texttt{Filter} is added to align with the update property (i.e. \texttt{FILTER(?seasons = 4)}). This incremental approach lowers the risk of introducing new errors, resulting in at most 3\% of queries getting a reduced F1 score after the self-refinement across all datasets and models.

    \item \textbf{Stage 4: Repeat if Necessary (line 11 back to line 2).}  
    The system executes the refined query anew, generating updated results and a refreshed NLE, then going back to stage (1) above. This loop repeats until satisfactory outputs are obtained (from either human or LLM feedback) or the process reaches a designated iteration limit. For \textbf{Raw Query}, the final feedback is \textit{``The query aligns well with the natural language question and produces accurate results.''}, indicating the refinement successfully finishes and the refined query is now aligned with the user's intention.
\end{enumerate}

By interweaving targeted feedback, query execution, and incremental corrections (including entity/property lookups), \our\  gradually rectifies any discrepancy between the user’s question and the evolving SPARQL query. Although this structure naturally accommodates actual user feedback during iteration, it is also possible to use LLM to simulate user input. We present this \textit{self-refinement} variant in Section~\ref{sec:self-ref}.

\subsubsection{Role and Significance of the NLE}
Although the NLE is generated or updated in stage~(1) (lines 3-6) of Algorithm~\ref{alg:interactive-refine}, it informs each iteration: \textbf{(a) Clarity for users:} By expressing the query’s triples, filters, or other clauses in a human-friendly style, the NLE allows both non-specialists and domain experts to pinpoint problematic regions needing attention; \textbf{(b) Anchor for feedback and tool calls:} The NLE offers a structured blueprint of the SPARQL statement, so the user/LLM can reference specific IRIs or variables before invoking the relevant lookup tool; \textbf{(c) Semantic continuity:} After every iteration, the NLE is updated to mirror the refined query, ensuring subsequent feedback remains accurate and consistent with the latest version. Overall, the NLE bridges the gap between original SPARQL code and high-level user reasoning, ensuring coherent and iterative query refinement, as empirically proven by experiments in Section~\ref{sec:eval}.

\subsubsection{Self-Refinement Baseline}\label{sec:self-ref}
Under normal circumstances, stage~(2) (line 7) of Algorithm~\ref{alg:interactive-refine} assumes that human users (or domain experts) review the query outputs and provide feedback on whether additional filters, entity substitutions, or property adjustments are needed. However, when real-time user involvement is unavailable or impractical, we employ a \textit{self-refinement} variant that demonstrates the workflow’s viability under reproducible conditions. In this mode, the LLM assumes both roles, i.e. \textit{feedback} and \textit{refine}, by: \textbf{(a)} generating feedback (the \texttt{getFeedback} function in stage (2)) based on discrepancies between the NLE, the executed query’s results, and the intended user question; \textbf{(b)} replacing or modifying specific query elements (in stage~(3)) according to the LLM’s own self-issued feedback, all while preserving validated segments from earlier iterations.


By embedding these automated feedback cycles and tool calls into the established refinement loop, \our\ confirms the framework’s capacity to converge on correct SPARQL queries without relying on direct human input; once user interaction becomes feasible (\textit{e.g.}, when a domain expert is available to oversee the refinement process), the system seamlessly transitions into a fully interactive mode, with the user or LLM calling upon the same entity/property search tools as needed. In both automated and human-in-the-loop modes, \our\ synthesizes feedback, applying on-demand search tools whenever domain knowledge is lacking, and documenting each refinement cycle through the NLE to maintain transparency about which parts of the query have been altered and why. This deliberate integration of query execution, feedback, and NLE documentation ensures reliable convergence (since flawed edits are promptly identified and revised) without sacrificing previously validated components. By streamlining evaluation in controlled environments and paving the way for a robust, flexible human-in-the-loop approach, this design provides a solid foundation for the subsequent evaluation of our interactive refinement methodology.



\section{Experimental Evaluation} \label{sec:eval}
The experiments are designed to assess \our's effectiveness by measuring two things: (1) the quality of the generated NLEs, and (2) the accuracy of the SPARQL queries produced by our interactive refinement pipeline. 

\subsection{Implementation}

We implemented \our\ in Python, ensuring compatibility with any operating system supporting Python 3.8 or higher. The code base comprises approximately 6k lines, spanning modules for data processing, NLE generation and refinement, interactive query refinement, and evaluation. \our\ is lightweight in terms of resource consumption and can be executed on a standard personal laptop. The only external requirements are either API access to proprietary models such as \gpto\ or \cld, or a GPU to serve open-source models like \qwt. The full end-to-end workflow for a single query incurs an average cost of \$0.03 when using \gpto, which can be reduced by 10–15$\times$ by switching to more efficient models like \gptom. When deploying open-source models locally, a single NVIDIA A100 GPU is sufficient to host LLMs with up to 32 billion parameters.
\subsection{Experimental Setup}

\paragraph{Datasets} We utilize a comprehensive set of SPARQL query benchmarks. The datasets employed are related to knowledge graphs Wikidata (\qaldt, \qaldnw), and DBpedia (\qaldnd). The Question Answering over Linked Data (\qald) series comprises human-annotated datasets designed to benchmark question-answering systems over linked data. \qaldn\ offers a good coverage of SPARQL queries on both knowledge graphs. \qaldt\ further advances \qaldn\ by increasing the complexity and size of the dataset, offering a more challenging benchmark for evaluating systems over Wikidata \cite{usbeck2023qald10}. Our evaluation framework ensures a robust assessment of \our\ across query complexity and natural language understanding.  We choose the human-annotated \qald\ datasets because they reflect practical usage scenarios. Note that we only consider \texttt{SELECT} and \texttt{ASK} queries as they are most frequently used as well as the only two types of SPARQL commands contained in all the human-labeled datasets.

\paragraph{LLMs and Query Engines} \label{subsec:llm_and_engine}

Our experimental framework is designed to be model-agnostic, allowing seamless integration of various LLMs that exhibit basic code understanding and instruction-following capabilities, although their performance may differ. We evaluate the performance of five LLMs: \gpto, \gptom, \cld, \qwt, and \qwf\footnote{The versions of proprietary models are: \gpto\ (gpt-4o-2024-08-06), \gptom\ (gpt-4o-mini-2024-07-18) and \cld\ (claude-3-5-sonnet-20241022).}. This versatility across proprietary and open-source LLMs enables us to interchange these models without compromising the integrity of our pipeline. We rely on the public SPARQL endpoints provided by DBpedia and Wikidata to execute SPARQL queries and retrieve data from knowledge graphs. The DBpedia SPARQL endpoint\footnote{\url{http://dbpedia.org/sparql}} facilitates direct querying of the DBpedia dataset.  The Wikidata SPARQL service\footnote{\url{https://query.wikidata.org/}} provides structured access to the Wikidata knowledge base. These endpoints are essential for obtaining the precise information required for our experimental evaluations.

\begin{table*}[ht]
    \centering
    \caption{Experimental Results on \qaldt, \qaldnw, and \qaldnd. (a) Raw Generation; (b) Upper Bound Generation. (c) Self-refine Generation. Macro-averaged F1 scores are applied here.}
    \label{tab:results}
    
    \begin{subtable}[t]{\textwidth}
        \centering
        \caption{Raw Generation}
        \small
        \begin{tabular}{lccc|ccc|ccc}
            \toprule
            \multirow{2}{*}{Model} & \multicolumn{3}{c}{\qaldt} & \multicolumn{3}{c}{\qaldnw} & \multicolumn{3}{c}{\qaldnd} \\
            \cmidrule(lr){2-4} \cmidrule(lr){5-7} \cmidrule(lr){8-10}
             & Prec. & Recall & F1 & Prec. & Recall & F1 & Prec. & Recall & F1 \\
            \midrule
            \gpto\       & 0.135  & 0.143  & 0.136  & 0.280  & 0.275  & 0.264  & 0.473  & 0.480  & 0.467 \\
 
            \gptom\  & 0.035  & 0.034 & 0.033 & 0.278  & 0.279  & 0.265  & 0.435  & 0.452  & 0.430 \\
            \cld\      & 0.172  & 0.176  & 0.172  & 0.356  & 0.376  & 0.350  & 0.524  & 0.547  & 0.523 \\
            \qwt\    & 0.019 & 0.021 & 0.020 & 0.023 & 0.036 & 0.026 & 0.314  & 0.316  & 0.310 \\

            \qwf\    & 0.057  & 0.057  & 0.057  & 0.009 & 0.015 & 0.010 & 0.315  & 0.325  & 0.312 \\
            \bottomrule
        \end{tabular}
    \end{subtable}
    
    \vspace{1em} 
    
    \begin{subtable}[t]{\textwidth}
        \centering
        \caption{Upper Bound Generation}
        \small
        \begin{tabular}{lccc|ccc|ccc}
            \toprule
            \multirow{2}{*}{Model} & \multicolumn{3}{c}{\qaldt} & \multicolumn{3}{c}{\qaldnw} & \multicolumn{3}{c}{\qaldnd} \\
            \cmidrule(lr){2-4} \cmidrule(lr){5-7} \cmidrule(lr){8-10}
             & Prec. & Recall & F1 & Prec. & Recall & F1 & Prec. & Recall & F1 \\
            \midrule
            \gpto\       & 0.930  & 0.930  & 0.930  & 0.833  & 0.831  & 0.837  & 0.931  & 0.931  & 0.931 \\
           \gptom\  & 0.948  & 0.947  & 0.947  & 0.838  & 0.835  & 0.836  & 0.938  & 0.937  & 0.937 \\
            \cld\      & 0.873  & 0.873  & 0.873  & 0.671  & 0.673  & 0.671  & 0.964  & 0.964  & 0.964 \\
            \qwt\   & 0.957  & 0.961  & 0.959  & 0.955  & 0.960  & 0.957  & 0.974  & 0.974  & 0.974 \\
            \qwf\    & 0.953  & 0.951  & 0.951  & 0.935  & 0.936  & 0.935  & 0.914  & 0.913  & 0.913 \\
            \bottomrule
        \end{tabular}
    \end{subtable}

    \begin{subtable}[t]{\textwidth}
        \centering
        \caption{Self-refine Generation}
        \small
        \begin{tabular}{lccc|ccc|ccc}
            \toprule
            \multirow{2}{*}{Model} & \multicolumn{3}{c}{\qaldt} & \multicolumn{3}{c}{\qaldnw} & \multicolumn{3}{c}{\qaldnd} \\
            \cmidrule(lr){2-4} \cmidrule(lr){5-7} \cmidrule(lr){8-10}
             & Prec. & Recall & F1 & Prec. & Recall & F1 & Prec. & Recall & F1 \\
            \midrule
           \gpto\   &  0.389   & 0.408  & 0.393  & 0.581 & 0.585  & 0.561  & 0.549  & 0.551  & 0.532 \\
            \gptom\  & 0.326  &  0.345 & 0.328  & 0.544  & 0.546  & 0.553  & 0.522  & 0.521  & 0.512 \\
            \cld\    & 0.377 & 0.408  & 0.383  & 0.567  & 0.588  &  0.560 & 0.574  & 0.593  & 0.567     \\
            \qwt\   & 0.314  & 0.361  & 0.325   &  0.382 & 0.497  & 0.411  & 0.532  & 0.530 & 0.523  \\
            \qwf\ & 0.259  & 0.291  & 0.266  & 0.326  & 0.361  & 0.328  & 0.443  &  0.452 &  0.439  \\
            \bottomrule
        \end{tabular}
    \end{subtable}
\end{table*}

\subsection{Accuracy of \our}
\label{subsec:eval-query-ref}

We evaluate \our\ by measuring how accurately it help align SPARQL queries with user questions. Two distinct experiments are conducted to evaluate \our\ against certain baselines. In the first, we examine an \textit{upper-bound scenario} where ground-truth NLEs are treated as fully correct and used to guide a single-pass SPARQL generation. In the second, we test a more \textit{realistic self-refinement scenario}, in which the system iteratively adjusts queries over several rounds, drawing on automatically generated NLEs. In both cases, the generated queries are executed and compared to the results of known ground-truth queries.


\subsubsection{Metrics and Setup.}
We first execute each query on the target knowledge graph (Wikidata or DBpedia) with the online API service (as mentioned in Section ~\ref{subsec:llm_and_engine}) and then compare its returned result set with that of the ground-truth query. For \texttt{ASK} queries, we directly check whether both queries produce the same boolean value. For \texttt{SELECT} queries, we gather and compare the sets of returned tuples. We then compute precision, recall, and F1 score to indicate how closely the results align. These metrics are collected across all queries tested, and we also aggregate them to identify overall performance and potential error patterns, i.e., average precision/recall and macro-averaged F1 (F1 scores mentioned in the following sections are macro-averaged F1 if not specified otherwise).

\subsubsection{Raw Generation and Upper-Bound Generation} \label{subsec:baseline-upperbound}

We compare \our\ against two contrasting scenarios. The results are shown in Table~\ref{tab:results}.

\paragraph{Raw generation.} In this setup, the LLM directly produces a SPARQL query from a user’s natural-language question, without any intermediate explanation or refinement. This represents our \textit{baseline}, testing how well the LLM can handle SPARQL generation in a single pass when guided only by a short prompt containing the natural language question and the knowledge graph it is based on. As Table~\ref{tab:results}(a) demonstrates, raw generation typically yields very low F1 scores. To better understand the reason of this poor performance, we take the 123 examples where our self-refinement loop ultimately improved the raw query and run a fine-grained keyword search over every feedback message flagged as an actual error (using GPT4o on \qaldt). We count all issues per entry (so one entry might contribute multiple error counts, at most one count for each type of error). The most frequent mistakes are incorrect entity IRIs (71) and incorrect property IRIs (55). Mistakes in all patterns other than BGPs appear 8 times (like \texttt{Filter} and \texttt{Bind}). We also observe 10 alignment‐to‐question errors and 8 execution result errors, which indicate the query does not match the user's question in general. In short, the raw general does not fail for lack of syntactic or logical SPARQL competence, but because the model cannot reliably select the right identifiers or fully understand the user’s question in a single pass.

\begin{table*}[ht]
  \centering
  \caption{Ablation study on design of NLE on \qaldt\ and \qaldnd\ datasets with both closed‑ and open‑sourced LLMs. Macro-averaged F1 scores are applied here.}
  \label{tab:nle_ablation}
  \setlength{\tabcolsep}{3pt}       
  \renewcommand{\arraystretch}{0.95}  
    \small

  \begin{tabular}{lccc|ccc|ccc|ccc}
    \toprule
    \multirow{2}{*}{Model \& Dataset}
      & \multicolumn{3}{c}{OD}
      & \multicolumn{3}{c}{BQB}
      & \multicolumn{3}{c}{NFS}
      & \multicolumn{3}{c}{BQFS} \\
    \cmidrule(lr){2-4}\cmidrule(lr){5-7}\cmidrule(lr){8-10}\cmidrule(lr){11-13}
      & Prec.\ & Recall & F1
      & Prec.\ & Recall & F1
      & Prec.\ & Recall & F1
      & Prec.\ & Recall & F1 \\
    \midrule
    \gpto\ + \qaldt   & 0.930  & 0.930  & 0.930 & 0.766 & 0.768 & 0.767 & 0.819 & 0.817 & 0.818 & 0.376 & 0.376 & 0.376 \\
    \qwt\ + \qaldt   & 0.957 & 0.961 & 0.959 & 0.693 & 0.697 & 0.693 & 0.881 & 0.889 & 0.883 & 0.404 & 0.404 & 0.403 \\
    \gpto\ + \qaldnd &   0.931  & 0.931  & 0.931  & 0.825  & 0.841  & 0.829  &  0.926 & 0.928 & 0.927 
 &  0.821  & 0.835   & 0.826 \\
    \qwt\ + \qaldnd  & 0.974 & 0.974 & 0.974 & 0.860 & 0.875 & 0.864 & 0.980 & 0.980 & 0.980 & 0.970 & 0.970 & 0.970 \\
    \bottomrule
  \end{tabular}
\end{table*}

\begin{table*}[ht]
  \centering
  \caption{Ablation study about the design of NLE on the performance of self-refinement on \qaldt\ and \qaldnd\ datasets with both closed‑ and open‑sourced LLMs. Macro-averaged F1 scores are applied here.}
  \label{tab:nle_ablation_self_refine}
  \setlength{\tabcolsep}{3pt}       
  \renewcommand{\arraystretch}{0.95}  
  \small

  \begin{tabular}{lccc|ccc|ccc|ccc}
    \toprule
    \multirow{2}{*}{Model \& Dataset}
      & \multicolumn{3}{c}{OD}
      & \multicolumn{3}{c}{BQB}
      & \multicolumn{3}{c}{NFS}
      & \multicolumn{3}{c}{BQFS} \\
    \cmidrule(lr){2-4}\cmidrule(lr){5-7}\cmidrule(lr){8-10}\cmidrule(lr){11-13}
      & Prec.\ & Recall & F1
      & Prec.\ & Recall & F1
      & Prec.\ & Recall & F1
      & Prec.\ & Recall & F1 \\
    \midrule
    \gpto\ + \qaldt  &  0.389   & 0.408  & 0.393 &  0.329 & 0.344 &  0.332 & 0.374 & 0.391 & 0.378 & 0.291 & 0.308 & 0.294 \\
    \qwt\ + \qaldt & 0.314  & 0.361  & 0.325   & 0.290  & 0.337 & 0.302 & 0.309  & 0.363 & 0.322 & 0.260 & 0.297 &  0.268\\
    \gpto\ + \qaldnd  &   0.549  & 0.551  & 0.532   &  0.545 & 0.533 & 0.521 & 0.543 & 0.538 & 0.525 & 0.521 & 0.515 & 0.503 \\
    \qwt\ + \qaldnd  &  0.532  & 0.530 & 0.523  & 0.498   & 0.501  & 0.491 & 0.488 & 0.486 & 0.480  & 0.515 & 0.514 & 0.508 \\
    \bottomrule
  \end{tabular}
\end{table*}

\paragraph{Upper-bound generation.} To establish a performance ceiling and assess how well a thorough NLE can guide query formulation, we test an \textit{upper-bound} scenario (Table~\ref{tab:results}(b)). Here, the LLM is given a \textit{ground truth} NLE generated by the ground truth query so it captures all relevant entities, properties, and structural details of the target SPARQL. It then translates this explanation into a final query in one step, which gets evaluated. As shown in Table~\ref{tab:results}(b), the generated queries achieve near-perfect precision, recall, and F1 score in this upper-bound scenario. This outcome underscores that, given a complete and accurate NLE, even one-step LLM-based query generation can align closely with the ground truth. Although this setting is impractical for routine deployment (because perfect NLEs are rarely available out of the box), it establishes an upper bound to judge how well \our\ performs. It also shows that, once the LLM has a well-structured explanation, it can generate near-flawless queries, even in a single pass. Crucially, this suggests \our\ can also serve as an effective ``handle'' for interactive refinement (Section~\ref{subsec:practical-self-ref}) if humans or automated feedback loops wish to iterate further.

\subsubsection{Practical Self-Refinement} \label{subsec:practical-self-ref}

We implement the self-refinement baseline that autonomously iterates up to five times to refine both the SPARQL query and its natural-language explanation based on detected errors. Table~\ref{tab:results}(c) shows that our \textbf{self-refinement} pipeline achieves substantially higher F1 scores than the baseline raw generation. Notably, no external agent (e.g., a human reviewer) provides feedback in this setup; the model itself identifies potential issues (such as incorrect IRIs or missing filters), invokes tool-based lookups when necessary, and revises the query accordingly. These results demonstrate that even a fully automated loop can meaningfully converge toward the user’s intended query. Furthermore, this setup also constitutes a \textit{good starting point for a human-in-the-loop} approach: once the model refines the query to a satisfactory baseline, a human expert (if available) can inspect and fine-tune it further, minimizing the manual workload required.

\subsection{Ablation Studies}

\subsubsection{Design Choice of NLE}

To evaluate the impact of our NLE design, we conduct two complementary experiments: a direct ablation study on the quality of generated queries (Table~\ref{tab:nle_ablation}) and a separate evaluation focusing on the effectiveness of the NLE during iterative self-refinement (Table~\ref{tab:nle_ablation_self_refine}). Both experiments compare four distinct configurations: \textbf{(a) Original Design (OD):} Incorporates structured semantic and hierarchical information extracted through rule-based methods from the AST, complemented by linguistically polished, LLM-refined narratives. Carefully designed few-shot examples in an accessible, structured format further guide the refinement; \textbf{(b) Bare Query Baseline (BQB):} Presents the original SPARQL query directly to the LLM, accompanied only by a generic instruction ("explain this query in natural language"), without structured guidance or few-shot examples; \textbf{(c) Bare Query with Few-Shots (BQFS):} Provides the original SPARQL query alongside structured few-shot examples derived from our methodology but omits explicit structured guidance from AST-derived rule-based NLEs, highlighting the influence of few-shot examples alone. \textbf{(d) No Few-Shots (NFS):} Employs the complete structured prompt but excludes few-shot examples, isolating the impact of explicit few-shot guidance.

As shown in Table~\ref{tab:nle_ablation}, our original NLE design provides superior results across both \qaldt\ and \qaldnd\ datasets when generating SPARQL queries directly from explanations. The OD approach achieves outstanding performance, reaching near-optimal F1 scores (0.977 for \gpto\ with \qaldt, 0.959 for \qwt\ with \qaldt, and 0.974 for \qwt\ with \qaldnd). These outcomes highlight that our structured explanations robustly preserve crucial query semantics and structural details, strongly aligning with insights from the human evaluation discussed in Section~\ref{subsec:human_res}.

The performance trend remains consistent in the iterative self-refinement scenario (Table~\ref{tab:nle_ablation_self_refine}). Although the absolute F1 scores are lower due to the complexity of iterative refinement without ground truth NLE, the OD approach continues to exhibit notable advantages. Specifically, \gpto\ combined with OD significantly outperforms all baseline configurations on both datasets. This indicates that our detailed and structured explanations provide crucial scaffolding for the LLM-driven self-refinement process, enabling more accurate and meaningful iterative improvements.

An additional noteworthy observation is that the NFS setting, despite omitting explicit structured few-shot guidance, retains complete query information, enabling LLMs to effectively interpret query details even when provided in less structured formats. Nevertheless, the structured few-shot examples significantly enhance interpretability and clarity, an effect further substantiated by human evaluation results in Sec.~\ref{subsec:human-nle-eval}.

Furthermore, simpler approaches such as BQB and BQFS occasionally achieve competitive performance, especially with the DBpedia dataset, benefiting from its semantic-rich identifiers and simpler query structures. Nevertheless, even in these favorable contexts, OD consistently surpasses other configurations, highlighting its robustness and consistent capability in capturing detailed query semantics.

Collectively, these experimental outcomes underscore the pivotal role of our structured and guided NLE design, both in direct SPARQL generation and iterative refinement, reinforcing its effectiveness and robustness.

\subsubsection{Design Choice Ablation Study for Self-Refinement}\label{subsec:abl-nle-design}

\begin{table*}[ht]
    \centering
    \caption{Ablation study on design of self-refinement with both closed- and open-sourced LLMs.  Macro-averaged F1 scores are applied here.}
    \label{tab:self_ref_ablation}
    \setlength{\tabcolsep}{3pt}       
    \renewcommand{\arraystretch}{0.95}  
    \small
    \begin{tabular}{lccc|ccc|ccc|ccc}
        \toprule
        \multirow{2}{*}{Model} & \multicolumn{3}{c}{Original} & \multicolumn{3}{c}{w/o NLE} & \multicolumn{3}{c}{w/o External Tools} & \multicolumn{3}{c}{w/o Both} \\
        \cmidrule(lr){2-4} \cmidrule(lr){5-7} \cmidrule(lr){8-10} \cmidrule(lr){11-13}
         & Prec. & Recall & F1 & Prec. & Recall & F1 & Prec. & Recall & F1 & Prec. & Recall & F1\\
        \midrule
        \gpto\ + \qaldt      &  0.389   & 0.408  & 0.393  & 0.347  & 0.370 &  0.351 & 0.112 & 0.123 & 0.112 & 0.086  & 0.098 & 0.086 \\
        \qwf\ + \qaldt    &  0.259  & 0.291  & 0.266  & 0.252 & 0.288 & 0.260 & 0.082  & 0.098  & 0.084 & 0.062  & 0.073  & 0.063 \\
        \qwt\ + \qaldt  & 0.314  & 0.361  & 0.325 & 0.289 & 0.336 & 0.300 & 0.080 & 0.092 & 0.083  & 0.044  & 0.047 & 0.045 \\
        \gpto\ + \qaldnd    &   0.549  & 0.551  & 0.532   & 0.536  & 0.536 & 0.518 &  0.496 & 0.506  & 0.490 & 0.501  & 0.514 & 0.497 \\
        \bottomrule
    \end{tabular}
\end{table*}

To quantify the impact of our NLE design and external tool integration within the self-refinement loop (cf. Section~\ref{sec:self-ref}), we perform an ablation study on both \qaldt\ and \qaldnd\ datasets. Table~\ref{tab:self_ref_ablation} summarizes the performance of representative models—\gpto, \qwf, and \qwt—under four distinct configurations: \textbf{(a) OD:} Implements the full self-refinement pipeline, integrating the two-step NLE (rule-based extraction followed by LLM refinement) and external entity/property search tools; \textbf{(b) w/o NLE:} Omits the NLE module, thus relying solely on bare feedback of the external tools without structured explanation; \textbf{(c) w/o External Tools:} Retains NLE guidance but excludes external tool calls for entity/property resolution and query execution; \textbf{(d) w/o Both:} Removes both the NLE and external tool components, forcing iterative refinement with minimal guidance.


For \gpto, the comprehensive original configuration achieves an F1 score of 0.402 on \qaldt, which declines moderately to 0.351 without the NLE component. The absence of external tools results in a drastic performance drop to 0.112, and removing both elements further diminishes performance to an F1 score of 0.086. A similar trend emerges with \qwf\ on  \qaldt\ and \gpto\ on \qaldt\ datasets.  These findings clearly highlight: \textbf{(a) critical role of NLE:} Structured explanations significantly enhance iterative refinement by providing interpretable query logic. The removal of NLE moderately decreases performance, indicating its vital role, especially in conjunction with external tools; \textbf{(b) essential external tools:} External entity/property resolution tools are crucial for accurately identifying IRIs and verifying query executions, evidenced by substantial performance drops upon their removal;  \textbf{(c) synergistic interaction:} Optimal outcomes occur when both NLE and external tools are integrated, confirming the structured NLE's role in guiding external tool invocation and ensuring robust query refinement.

\subsubsection{Design Choice of Maximum Refinement Iteration in Self‐Refinement}
To quantify how the maximum number of self‐refinement iterations ($N$ in Algorithm~\ref{alg:interactive-refine}) affects final query accuracy, we evaluate \gpto\ and \qwt\ on \qaldt\ with $N\in\{1,\,5,\,10\}$ ($N=5$ is our default). At each setting, the loop halts early if the generated SPARQL matches ground truth; otherwise, it proceeds until reaching $N$. We report precision, recall, and macro‐averaged F1 score of the final query.

\begin{table}[ht]
\centering
\footnotesize
\caption{Ablation on maximum number of self‐refinement iterations ($N$) for QALD‐10.}
\label{tab:iteration_ablation}
\setlength{\tabcolsep}{3pt}
\small
\begin{tabular}{l c c c c}
\toprule
\textbf{Model}            & \textbf{$N$} & \textbf{Precision} & \textbf{Recall} & \textbf{F1} \\ 
\midrule
\multirow{3}{*}{GPT‐4o}   & 1  & 0.303 & 0.326 & 0.307 \\ 
                          & 5  &  0.389   & 0.408  & 0.393 \\ 
                          & 10 & 0.410 & 0.442 & 0.417 \\ 
\midrule
\multirow{3}{*}{Qwen‐2.5‐32B} & 1  & 0.220 & 0.245 & 0.225 \\ 
                             & 5  & 0.314  & 0.361  & 0.325  \\ 
                             & 10 & 0.340 & 0.388 & 0.352 \\ 
\bottomrule
\end{tabular}
\vspace{-1em}
\end{table}

As Table~\ref{tab:iteration_ablation} shows, allowing only a single iteration ($N=1$) yields modest improvements: GPT‐4o achieves an F1 of 0.307, and Qwen‐2.5‐32B reaches 0.225, since one round of self‐refinement can correct simple errors but often leaves deeper misalignments (e.g., incorrect property IRIs) unresolved. When $N=5$, both models experience a substantial boost raising GPT‐4o’s F1 to 0.402 and Qwen‐2.5‐32B’s to 0.329, as a result of the fact that most queries converge within five rounds of lookup and minor structural edits. Extending to $N=10$ provides only marginal gains, with GPT‐4o’s F1 moving to 0.417 and Qwen‐2.5‐32B’s to 0.352, because the majority of self‐refinements have already converged by iteration five. These results confirm that while increasing $N$ from 1 to 5 is crucial for achieving high accuracy, doubling from 5 to 10 incurs diminishing returns and typically does not justify the extra computational cost.

\subsection{Human Evaluation of NLE Quality}
\label{subsec:human-nle-eval}

Complementing the benchmark evaluations, we conduct a comprehensive human evaluation to assess the perceived quality of our NLEs directly. This evaluation involves a head-to-head comparative study among the four ablation configurations detailed previously in Section~\ref{subsec:abl-nle-design}: \textbf{OD}, \textbf{BQB}, \textbf{BQFS}, and \textbf{NFS}. 

\subsubsection{Evaluation Setup} \label{subsec:nle_human_setting}
The study focuses on essential dimensions, including dataset complexity and variations in LLMs.

\paragraph{Datasets and Knowledge Graphs \& LLM Variants.} We evaluate explanations generated for two datasets of differing complexity: \qaldt\ (higher complexity) and \qaldnd\ (lower complexity). This allows us to examine how the complexity of datasets and the underlying knowledge graphs (DBpedia versus Wikidata) influence explanation quality. We evaluate with both closed-source models (e.g., \gpto) and open-source models (e.g., \qwt) to confirm the general applicability and robustness of our NLE approach across different LLM architectures.


\paragraph{Experimental Design.} Each participant sees a series of head-to-head comparisons. In each comparison, they are shown two NLEs side by side for the same SPARQL query: one produced by the Original Design (OD) and the other by exactly one of the three ablated configurations (BQB, BQFS, or NFS). The participants are not told which one is from which system. For every pair, participants provide four independent dimension scores on a five-point Likert scale (1 = Very Poor, 5 = Very Good) and then indicate an overall preference (OD wins, OD loses, or Tie). The four dimensions are: \textbf{(a) Aesthetics:} Readability, highlighting, formatting, and overall presentation structure; \textbf{(b) Clarity:} Use of clear, everyday language, logical flow, and ease of understanding the query’s intent; \textbf{(c) Completeness:} Coverage of all critical SPARQL components—variables, graph patterns, filters, subqueries, prefixes, and modifiers. \textbf{(d) Usefulness:} Whether the explanation helps a reader unfamiliar with the query to understand, debug, extend, or modify it accurately.
Although participants often rely on their dimension scores when choosing a winner, the overall preference is recorded separately to capture their holistic judgment. A detailed protocol of the human study given to the participants can be found in Section 2 of Appendix. 

\subsubsection{Participant Task and Scoring Criteria} \label{subsec:participant-task}

\begin{figure}
  \centering
  \begin{subfigure}[b]{0.48\textwidth}
    \includegraphics[width=\textwidth]{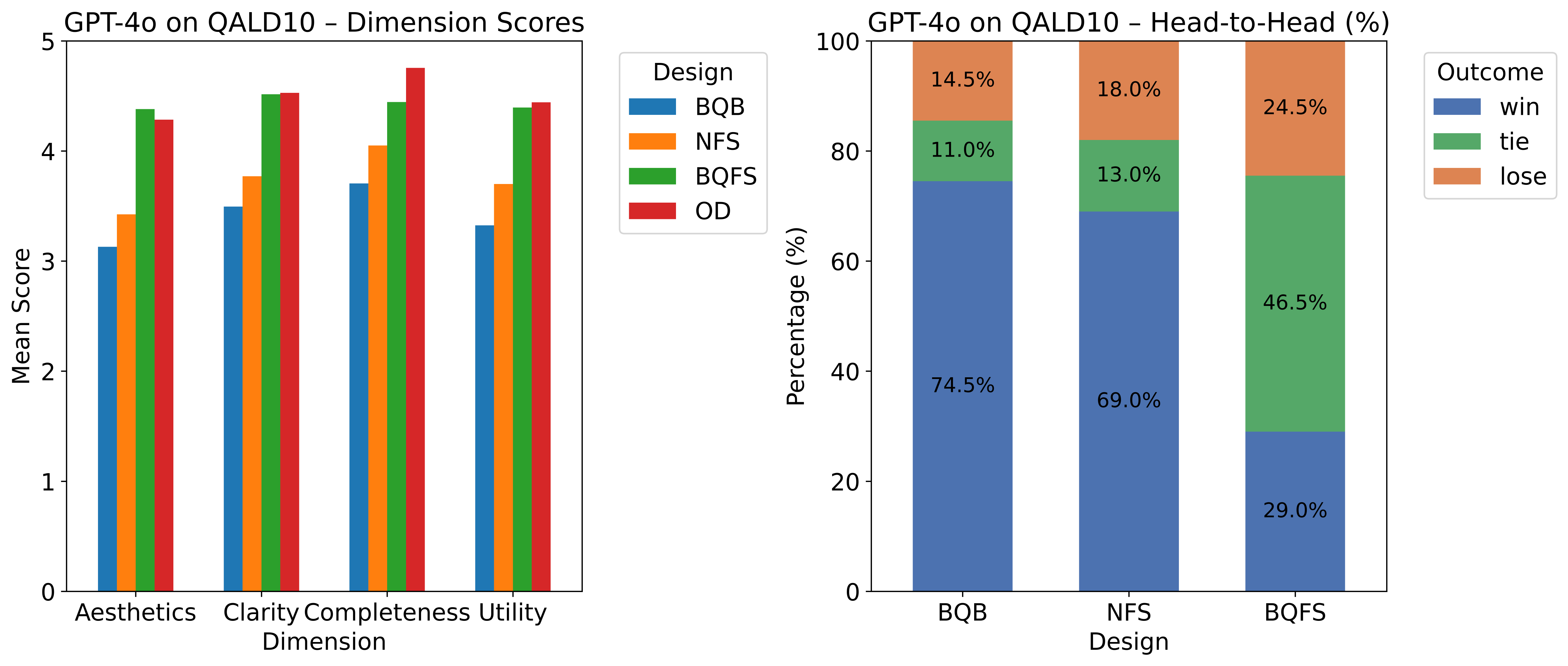}
    \caption{GPT-4o on QALD10}
    \label{fig:human1}
  \end{subfigure}
  \hfill
  \begin{subfigure}[b]{0.48\textwidth}
    \includegraphics[width=\textwidth]{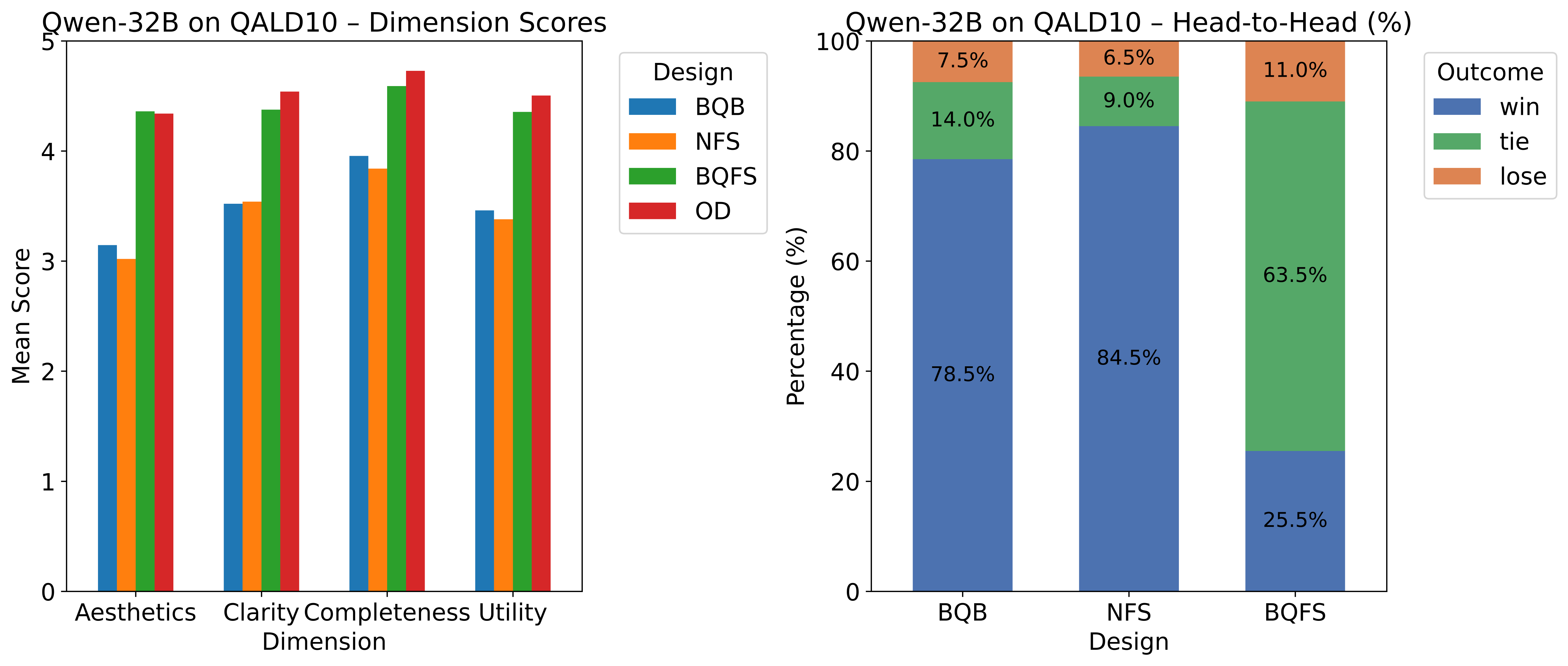}
    \caption{Qwen-32B on QALD10}
    \label{fig:human2}
  \end{subfigure}
  \hfill
  \begin{subfigure}[b]{0.48\textwidth}
    \includegraphics[width=\textwidth]{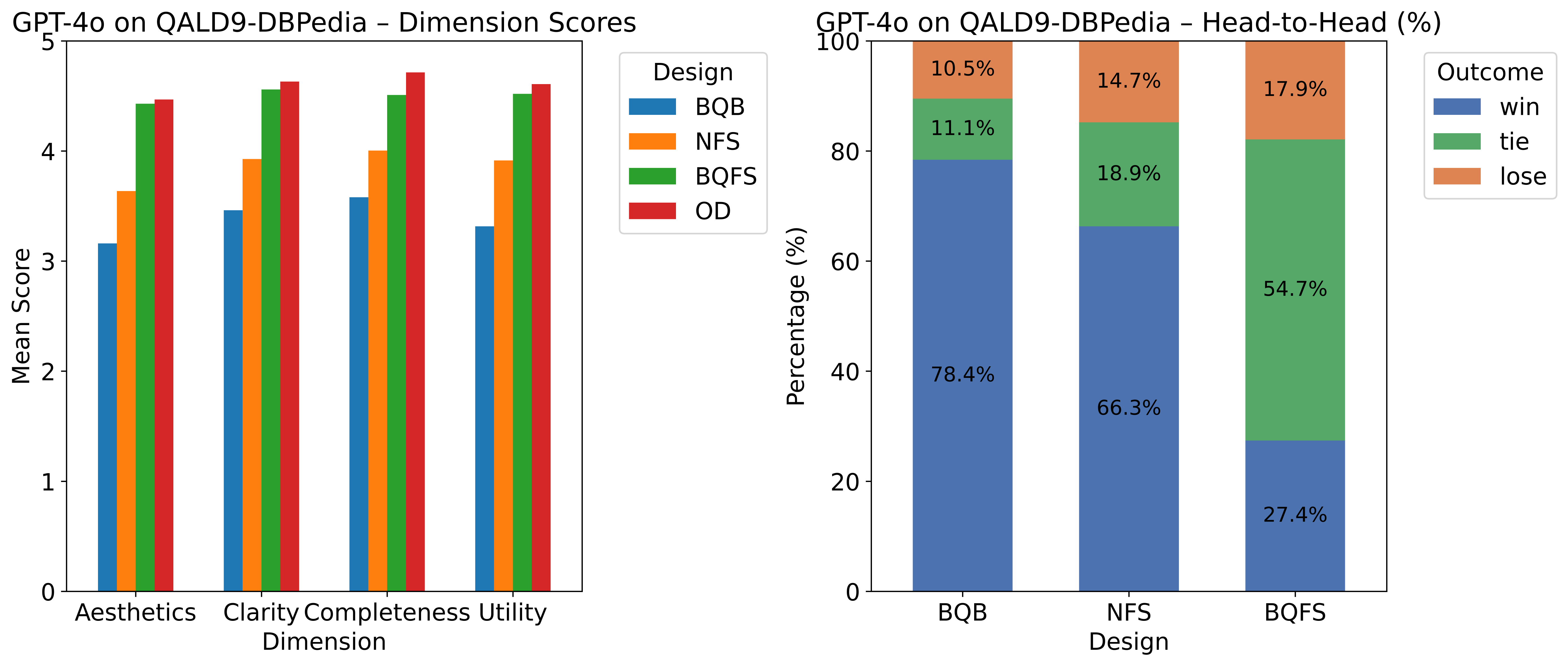}
    \caption{GPT-4o on QALD9-DBPedia}
    \label{fig:human3}
  \end{subfigure}
  \caption{Human evaluation results for the three conditions. Each subfigure shows mean dimension scores (left) and head-to-head percentages (right) for OD, BQB, NFS and BQFS.}
  \label{fig:human_evaluation} 
  \vspace{-1.em}
\end{figure}

We recruit graduate-level participants with basic familiarity with knowledge graphs and databases but varied expertise with SPARQL to ensure broad, representative feedback. The experiment involves 20 participants with a gender distribution of 8 females and 12 males. The participants are natively from 7 countries. 

\subsubsection{Results and Analysis} \label{subsec:human_res}
The human evaluation results, illustrated in Figure~\ref{fig:human_evaluation}, highlight the consistent advantage of our OD across all assessed dimensions and comparison settings. Participants consistently rate OD highest for completeness and usefulness across both models (GPT-4o and Qwen-2.5-32B) and datasets (QALD-9 DBpedia and QALD-10 Wikidata). This clearly underscores the effectiveness of integrating structured semantic extraction from SPARQL queries with carefully designed few-shot examples.

Among the ablation configurations, BQFS was appreciated for its intuitive formatting, yielding high aesthetic and clarity scores. Nonetheless, participants consistently identified crucial semantic details missing from BQFS explanations, underscoring that visual improvements alone cannot substitute for structured semantic extraction. NFS explanations retained comprehensive semantic content but lacked intuitive formatting. While recognized for completeness, these explanations were seen as less readable and somewhat mechanical, reinforcing the necessity of structured examples for user-friendliness. BQB consistently received the lowest ratings due to the absence of both structured content and formatting guidance. Participants frequently noted its lack of clarity, incomplete explanations, and poor utility for practical use.

In terms of overall preference, OD was strongly favored over both BQB and NFS. Preferences between OD and BQFS were more mixed, with participants acknowledging BQFS’s appealing format, yet consistently preferring OD for its comprehensive content coverage.

These results highlight that effectively combining structured semantic extraction with intuitive, example-based formatting is essential for producing high-quality natural language explanations for SPARQL queries.
\section{Related Work} \label{sec:related}

Various approaches have been proposed to address the challenges users face when interacting with RDF data with or without SPARQL. We review them in this section.

\subsection{One-turn NL2SPARQL Generation} \label{subsec:sparql_gen}

Recent advances in one-turn SPARQL query generation from natural language, particularly with the integration of LLMs, have significantly improved the efficiency and broadened the functionality of transforming natural language questions into SPARQL queries. Bustamante and Takeda ~\cite{bustamante2024sparqlgenerationentitypretrained} enhance SPARQL generation by pre-training GPT models on entities, improving entity linking and query translation accuracy. Banerjee et al.~\cite{banerjee2023roleoutputvocabularyt2t} demonstrate the importance of tailoring output vocabularies in text-to-text models, achieving significant performance gains. The SPARKLE framework \cite{lee2024sparkleenhancingsparqlgeneration} integrates knowledge graph information directly into the decoding process, aligning natural language inputs with SPARQL query structures. Furthermore, fine-tuning models like OpenLLaMA for domain-specific knowledge graphs, such as those in life sciences \cite{rangel2024sparqlgenerationanalysisfinetuning}, have been shown to improve query accuracy in specialized fields. The SPINACH framework \cite{liu2024spinachsparqlbasedinformationnavigation} enhances information navigation capabilities for complex real-world questions through SPARQL-based methods. ArcaneQA \cite{gu2022arcaneqadynamicprograminduction} leverages dynamic program induction and contextualized encoding to enhance knowledge base question answering. Lastly, few-shot learning approaches \cite{li-etal-2023-shot} demonstrate the potential of LLMs to generate accurate SPARQL queries with minimal training data. While works have improved query generation capabilities, they still face challenges in delivering consistent accuracy, often failing to fully capture the user's intent or retrieve precise results.

\subsection{Interactive SPARQL Query Refinement } \label{subsec:interactive_sparql}

Interactive query refinement has garnered some attention, though it remains relatively underexplored. SPARQLIt \cite{icde21_9458607} is a tool that assists users in interactively refining SPARQL queries. Users can iteratively refine their queries based on intermediate results, enabling a more user-friendly and precise query formulation process. Similarly, Abramovitz et al.~\cite{icde18_8509280} introduce an interactive inference approach that utilizes provenance information to enhance the accuracy and relevance of SPARQL query results. The PAROT framework \cite{OCHIENG2020100024} translates natural language queries into SPARQL using dependency-based heuristics, enabling users to handle complex queries more effectively. Jian et al.~\cite{sigmod20_JianWLZC20} study the theoretical complexity of query modification through restriction and relaxation, modeling it as a formal optimization problem. However, their work remains largely abstract and does not address interactive usability or user interpretation. Despite their advances, these systems often fail to produce human-friendly explanations systematically, a gap that our method fills by integrating structured NLEs.

\subsection{Knowledge-Based Question-Answering Platforms} \label{subsec:kbqa}

Unlike directly generating SPARQL queries, several studies have explored universal question-answering platforms \cite{Bouziane:2015aa} and their application to knowledge graphs. Omar et al. discuss the development of a universal QA platform that integrates various techniques to efficiently answer questions on knowledge graphs \cite{sigmod23_Omar:2023aa}. Similarly, StructGPT \cite{jiang2023structgptgeneralframeworklarge} introduces a framework that enables LLMs to reason over structured data like knowledge graphs and databases, enhancing their QA capabilities. Recent advancements in few-shot and multitask learning have also contributed to the field. Li et al.~\cite{li-etal-2023-shot} explore few-shot in-context learning for knowledge base question answering, aiming to improve performance with minimal training data. The UnifiedSKG framework \cite{xie-etal-2022-unifiedskg} unifies 21 structured knowledge grounding tasks into a text-to-text format, facilitating multi-task learning and improving model performance across various tasks. While KBQA systems effectively return answers, they generally lack transparency regarding the underlying query logic, which is a limitation that prevents users from verifying results, correcting errors, or learning how the answer was derived. Our work overcomes this gap by supplying explicit NLEs that expose the full SPARQL reasoning chain.

\section{Conclusion} \label{sec:conclusion}
We presented \our, interactive SPARQL query generation and refinement system that unifies a rule-based AST-driven explanation phase with LLM-based refinement. This two-stage pipeline significantly lowers the barrier to SPARQL formulation by offering structured NLEs, along with iterative self-refinement or user feedback to maintain alignment with evolving query goals. Evaluations on \qald\ benchmarks show \our\ outperforming baseline methods in accuracy, refinement capability, and user satisfaction; human studies further confirm improvements in explanation clarity, aesthetic, utility, and completeness.

In the future, we plan to expand \our\ to perform even better on advanced SPARQL features like nested queries and more intricate property paths, while refining entity and property linking for specialized knowledge bases. Deeper user-LLM collaboration models may also yield more robust and accurate query formulation. We believe these directions will help foster a more transparent, iterative SPARQL environment, ultimately empowering both novice and expert users to craft sophisticated, precise queries.


\bibliography{publications,sample-base}

\begin{thebibliography}{45}
\providecommand{\natexlab}[1]{#1}

\bibitem[{Abdelaziz et~al.(2017)Abdelaziz, Harbi, Khayyat, and Kalnis}]{vldb17_AbdelazizHKK17}
Ibrahim Abdelaziz, Razen Harbi, Zuhair Khayyat, and Panos Kalnis. 2017.
\newblock A survey and experimental comparison of distributed {SPARQL} engines for very large {RDF} data.
\newblock \emph{Proc. VLDB Endowment}, 10(13):2049--2060.

\bibitem[{Abramovitz et~al.(2018)Abramovitz, Deutch, and Gilad}]{icde18_8509280}
Efrat Abramovitz, Daniel Deutch, and Amir Gilad. 2018.
\newblock \href {https://doi.org/10.1109/ICDE.2018.00059} {Interactive inference of sparql queries using provenance}.
\newblock In \emph{Proc. 34th IEEE Int. Conf. on Data Engineering}, pages 581--592.

\bibitem[{Ali et~al.(2022)Ali, Saleem, Yao, Hogan, and Ngomo}]{Ali:2022vx}
Waqas Ali, Muhammad Saleem, Bin Yao, Aidan Hogan, and Axel-Cyrille~Ngonga Ngomo. 2022.
\newblock \href {https://doi.org/10.1007/s00778-021-00711-3} {A survey of {RDF} stores \& {SPARQL} engines for querying knowledge graphs}.
\newblock \emph{VLDB J.}, 31(3):1--26.

\bibitem[{Amsterdamer and Callen(2021)}]{icde21_9458607}
Yael Amsterdamer and Yehuda Callen. 2021.
\newblock \href {https://doi.org/10.1109/ICDE51399.2021.00295} {Sparqlit: Interactive sparql query refinement}.
\newblock In \emph{Proc. 37th IEEE Int. Conf. on Data Engineering}, pages 2649--2652.

\bibitem[{Angles et~al.(2022)Angles, Aranda, Hogan, Rojas, and Vrgo{\v{c}}}]{Angles:2022aa}
Renzo Angles, Carlos~Buil Aranda, Aidan Hogan, Carlos Rojas, and Domagoj Vrgo{\v{c}}. 2022.
\newblock \href {https://link.springer.com/chapter/10.1007/978-3-031-19433-7_41} {Wdbench: A wikidata graph query benchmark}.
\newblock In \emph{Proc. 21st Int. Semantic Web Conf.}, pages 714--731, Cham. Springer International Publishing.

\bibitem[{Angles and Gutierrez(2008)}]{angles:2008aa}
Renzo Angles and Claudio Gutierrez. 2008.
\newblock The expressive power of {SPARQL}.
\newblock In \emph{Proc. 7th Int. Semantic Web Conf.}, pages 114--129.

\bibitem[{Arenas and P{\'e}rez(2011)}]{pods11_arenasp11}
Marcelo Arenas and Jorge P{\'e}rez. 2011.
\newblock Querying semantic web data with {SPARQL}.
\newblock In \emph{Proc. 30th ACM SIGACT-SIGMOD-SIGART Symp. on Principles of Database Systems}, pages 305--316.

\bibitem[{Arenas and Ugarte(2017)}]{Arenas:2017aa}
Marcelo Arenas and Martin Ugarte. 2017.
\newblock \href {https://doi.org/10.1145/3129247} {Designing a query language for rdf: Marrying open and closed worlds}.
\newblock \emph{ACM Trans. Database Syst.}, 42(4):21:1--21:46.

\bibitem[{Arias et~al.(2011)Arias, Fern{\'a}ndez, Mart\'{\i}nez-Prieto, and de~la Fuente}]{ariasusewod2011}
Mario Arias, Javier~D. Fern{\'a}ndez, Miguel~A. Mart\'{\i}nez-Prieto, and Pablo de~la Fuente. 2011.
\newblock An empirical study of real-world {SPARQL} queries.
\newblock \emph{CoRR}, abs/1103.5043.

\bibitem[{Banerjee et~al.(2023)Banerjee, Nair, Usbeck, and Biemann}]{banerjee2023roleoutputvocabularyt2t}
Debayan Banerjee, Pranav~Ajit Nair, Ricardo Usbeck, and Chris Biemann. 2023.
\newblock \href {https://arxiv.org/abs/2305.15108} {The role of output vocabulary in t2t lms for sparql semantic parsing}.
\newblock \emph{Preprint}, arXiv:2305.15108.

\bibitem[{{Bio2RDF Project}()}]{bio2rdf}
{Bio2RDF Project}.
\newblock Bio2rdf: Linked data for the life sciences.
\newblock \url{https://bio2rdf.github.io/}.
\newblock Accessed: 2025-06-09.

\bibitem[{Bonifati et~al.(2017)Bonifati, Martens, and Timm}]{vldb18_BonifatiMT17}
Angela Bonifati, Wim Martens, and Thomas Timm. 2017.
\newblock An analytical study of large {SPARQL} query logs.
\newblock \emph{Proc. VLDB Endowment}, 11(2):149--161.

\bibitem[{Bonifati et~al.(2019)Bonifati, Martens, and Timm}]{www19_Bonifati:2019aa}
Angela Bonifati, Wim Martens, and Thomas Timm. 2019.
\newblock \href {https://doi.org/10.1145/3308558.3313472} {Navigating the maze of wikidata query logs}.
\newblock In \emph{Proc. 28th Int. World Wide Web Conf.}, pages 127--138.

\bibitem[{Bouziane et~al.(2015)Bouziane, Bouchiha, Doumi, and Malki}]{Bouziane:2015aa}
Abdelghani Bouziane, Djelloul Bouchiha, Noureddine Doumi, and Mimoun Malki. 2015.
\newblock \href {https://doi.org/10.1016/j.procs.2015.12.005} {Question answering systems: Survey and trends}.
\newblock \emph{Procedia Computer Science}, 73:366 -- 375.

\bibitem[{Bustamante and Takeda(2024)}]{bustamante2024sparqlgenerationentitypretrained}
Diego Bustamante and Hideaki Takeda. 2024.
\newblock \href {https://arxiv.org/abs/2402.00969} {Sparql generation with entity pre-trained gpt for kg question answering}.
\newblock \emph{Preprint}, arXiv:2402.00969.

\bibitem[{Cohen and Kim(2013)}]{cohen-kim-2013-evaluation}
K.~Bretonnel Cohen and Jin-Dong Kim. 2013.
\newblock \href {https://aclanthology.org/W13-5202/} {Evaluation of {SPARQL} query generation from natural language questions}.
\newblock In \emph{Proceedings of the Joint Workshop on {NLP}{\&}{LOD} and {SWAIE}: Semantic Web, Linked Open Data and Information Extraction}, pages 3--7, Hissar, Bulgaria. INCOMA Ltd. Shoumen, BULGARIA.

\bibitem[{Diallo et~al.(2024)Diallo, Reyd, and Zouaq}]{10662970}
Papa Abdou Karim~Karou Diallo, Samuel Reyd, and Amal Zouaq. 2024.
\newblock \href {https://doi.org/10.1109/ACCESS.2024.3453215} {A comprehensive evaluation of neural sparql query generation from natural language questions}.
\newblock \emph{IEEE Access}, 12:125057--125078.

\bibitem[{Diaz et~al.(2016)Diaz, Arenas, and Benedikt}]{vldb16_Diaz:2016}
Gonzalo~I. Diaz, Marcelo Arenas, and Michael Benedikt. 2016.
\newblock \href {https://doi.org/10.14778/3007263.3007302} {Sparqlbye: Querying {RDF} data by example}.
\newblock \emph{Proc. VLDB Endowment}, 9(13):1533--1536.

\bibitem[{Group(2013)}]{sparql11-overview}
W3C SPARQL~Working Group. 2013.
\newblock {SPARQL 1.1 Overview}.
\newblock \url{https://www.w3.org/TR/sparql11-overview/}.
\newblock Accessed: 2024-09-04.

\bibitem[{Gu and Su(2022)}]{gu2022arcaneqadynamicprograminduction}
Yu~Gu and Yu~Su. 2022.
\newblock \href {https://arxiv.org/abs/2204.08109} {Arcaneqa: Dynamic program induction and contextualized encoding for knowledge base question answering}.
\newblock \emph{Preprint}, arXiv:2204.08109.

\bibitem[{Harris and Seaborne(2013)}]{harris:aa}
Steve Harris and Andy Seaborne. 2013.
\newblock {SPARQL} 1.1 query language.
\newblock Accessible at http://www.w3.org/TR/sparql11-query/.
\newblock Last accessed November 2015.

\bibitem[{Hartig(2012)}]{esws12_hartig12}
Olaf Hartig. 2012.
\newblock \href {https://doi.org/10.1007/978-3-642-30284-8_8} {{SPARQL} for a web of linked data: Semantics and computability.}
\newblock In \emph{Proc. 9th Extended Semantic Web Conf.}, pages 8--23.

\bibitem[{Hartig et~al.(2009)Hartig, Bizer, and Freytag}]{hartig2009a}
Olaf Hartig, Christian Bizer, and J.C. Freytag. 2009.
\newblock Executing {SPARQL} queries over the web of linked data.
\newblock In \emph{Proc. 8th Int. Semantic Web Conf.}, pages 293--309.

\bibitem[{Helal et~al.(2021)Helal, Helali, Ammar, and Mansour}]{vldb21_Helal:2021}
Ahmed Helal, Mossad Helali, Khaled Ammar, and Essam Mansour. 2021.
\newblock \href {https://doi.org/10.14778/3476311.3476317} {A demonstration of kglac: A data discovery and enrichment platform for data science}.
\newblock \emph{Proc. VLDB Endowment}, 14(12):2675--2678.

\bibitem[{Jian et~al.(2020)Jian, Wang, Lei, Zheng, and Chen}]{sigmod20_JianWLZC20}
Xun Jian, Yue Wang, Xiayu Lei, Libin Zheng, and Lei Chen. 2020.
\newblock \href {https://doi.org/10.1145/3318464.3389695} {{SPARQL} rewriting: Towards desired results}.
\newblock In \emph{Proc. ACM SIGMOD Int. Conf. on Management of Data}, pages 1979--1993.

\bibitem[{Jiang et~al.(2023)Jiang, Zhou, Dong, Ye, Zhao, and Wen}]{jiang2023structgptgeneralframeworklarge}
Jinhao Jiang, Kun Zhou, Zican Dong, Keming Ye, Wayne~Xin Zhao, and Ji-Rong Wen. 2023.
\newblock \href {https://arxiv.org/abs/2305.09645} {Structgpt: A general framework for large language model to reason over structured data}.
\newblock \emph{Preprint}, arXiv:2305.09645.

\bibitem[{Lee and Shin(2024)}]{lee2024sparkleenhancingsparqlgeneration}
Jaebok Lee and Hyeonjeong Shin. 2024.
\newblock \href {https://arxiv.org/abs/2407.01626} {Sparkle: Enhancing sparql generation with direct kg integration in decoding}.
\newblock \emph{Preprint}, arXiv:2407.01626.

\bibitem[{Letelier et~al.(2012)Letelier, P{\'{e}}rez, Pichler, and Skritek}]{vldb12_LetelierPPS12}
Andr{\'{e}}s Letelier, Jorge P{\'{e}}rez, Reinhard Pichler, and Sebastian Skritek. 2012.
\newblock \href {https://doi.org/10.14778/2367502.2367547} {{SPAM:} {A} {SPARQL} analysis and manipulation tool}.
\newblock \emph{Proc. VLDB Endowment}, 5(12):1958--1961.

\bibitem[{Li et~al.(2023)Li, Ma, Zhuang, Gu, Su, and Chen}]{li-etal-2023-shot}
Tianle Li, Xueguang Ma, Alex Zhuang, Yu~Gu, Yu~Su, and Wenhu Chen. 2023.
\newblock \href {https://doi.org/10.18653/v1/2023.acl-long.385} {Few-shot in-context learning on knowledge base question answering}.
\newblock In \emph{Proceedings of the 61st Annual Meeting of the Association for Computational Linguistics (Volume 1: Long Papers)}, pages 6966--6980, Toronto, Canada. Association for Computational Linguistics.

\bibitem[{Liu et~al.(2021)Liu, Wang, Liu, Li, Fu, and Chai}]{icde21_9458632}
Baozhu Liu, Xin Wang, Pengkai Liu, Sizhuo Li, Qiang Fu, and Yunpeng Chai. 2021.
\newblock \href {https://doi.org/10.1109/ICDE51399.2021.00303} {Unikg: A unified interoperable knowledge graph database system}.
\newblock In \emph{Proc. 37th IEEE Int. Conf. on Data Engineering}, pages 2681--2684.

\bibitem[{Liu et~al.(2024)Liu, Semnani, Triedman, Xu, Zhao, and Lam}]{liu2024spinachsparqlbasedinformationnavigation}
Shicheng Liu, Sina~J. Semnani, Harold Triedman, Jialiang Xu, Isaac~Dan Zhao, and Monica~S. Lam. 2024.
\newblock \href {https://arxiv.org/abs/2407.11417} {Spinach: Sparql-based information navigation for challenging real-world questions}.
\newblock \emph{Preprint}, arXiv:2407.11417.

\bibitem[{Mohamed et~al.(2022)Mohamed, Abuoda, Ghanem, Kaoudi, and Aboulnaga}]{Mohamed:2022vu}
Aisha Mohamed, Ghadeer Abuoda, Abdurrahman Ghanem, Zoi Kaoudi, and Ashraf Aboulnaga. 2022.
\newblock \href {https://doi.org/10.1007/s00778-021-00690-5} {{RDFFrames}: Knowledge graph access for machine learning tools}.
\newblock \emph{VLDB J.}, 31(2):321--346.

\bibitem[{Morsey et~al.(2011)Morsey, Lehmann, Auer, and Ngomo}]{morsey:2011vn}
Mohamed Morsey, Jens Lehmann, S{\"o}ren Auer, and Axel-Cyrille~Ngonga Ngomo. 2011.
\newblock {DBpedia SPARQL} benchmark--performance assessment with real queries on real data.
\newblock In \emph{Proc. 10th Int. Semantic Web Conf.}, pages 454--469.

\bibitem[{Ochieng(2020)}]{OCHIENG2020100024}
Peter Ochieng. 2020.
\newblock \href {https://doi.org/10.1016/j.eswax.2020.100024} {Parot: Translating natural language to sparql}.
\newblock \emph{Expert Systems with Applications: X}, 5:100024.

\bibitem[{Omar et~al.(2023)Omar, Dhall, Kalnis, and Mansour}]{sigmod23_Omar:2023aa}
Reham Omar, Ishika Dhall, Panos Kalnis, and Essam Mansour. 2023.
\newblock \href {https://doi.org/10.1145/3588911} {A universal question-answering platform for knowledge graphs}.
\newblock \emph{Proc. ACM Manag. Data}, 1(1).

\bibitem[{P\'{e}rez et~al.(2009)P\'{e}rez, Arenas, and Gutierrez}]{perez2009}
Jorge P\'{e}rez, Marcelo Arenas, and Claudio Gutierrez. 2009.
\newblock Semantics and complexity of {SPARQL}.
\newblock \emph{ACM Trans. Database Syst.}, 34(3):1--45.

\bibitem[{Rangel et~al.(2024)Rangel, de~Farias, Sima, and Kobayashi}]{rangel2024sparqlgenerationanalysisfinetuning}
Julio~C. Rangel, Tarcisio~Mendes de~Farias, Ana~Claudia Sima, and Norio Kobayashi. 2024.
\newblock \href {https://arxiv.org/abs/2402.04627} {Sparql generation: an analysis on fine-tuning openllama for question answering over a life science knowledge graph}.
\newblock \emph{Preprint}, arXiv:2402.04627.

\bibitem[{Rony et~al.(2022)Rony, Kumar, Teucher, Kovriguina, and Lehmann}]{SGPT}
Md~Rashad Al~Hasan Rony, Uttam Kumar, Roman Teucher, Liubov Kovriguina, and Jens Lehmann. 2022.
\newblock \href {https://doi.org/10.1109/ACCESS.2022.3188714} {Sgpt: A generative approach for sparql query generation from natural language questions}.
\newblock \emph{IEEE Access}, 10:70712--70723.

\bibitem[{Shen et~al.(2015)Shen, Zou, \"{O}zsu, Chen, Li, Han, and Zhao}]{icde15_7113413}
X.~Shen, L.~Zou, M.~T. \"{O}zsu, L.~Chen, Y.~Li, S.~Han, and D.~Zhao. 2015.
\newblock \href {https://doi.org/10.1109/ICDE.2015.7113413} {A graph-based {RDF} triple store}.
\newblock In \emph{Proc. 31st IEEE Int. Conf. on Data Engineering}, pages 1508--1511.
\newblock System demonstration paper.

\bibitem[{Usbeck et~al.(2023)Usbeck, Yan, Perevalov, Jiang, Schulz, Kraft, Moeller, Huang, Reineke, Ngonga~Ngomo, Saleem, and Both}]{usbeck2023qald10}
Ricardo Usbeck, Xi~Yan, Aleksandr Perevalov, Longquan Jiang, Julius Schulz, Angelie Kraft, Cedric Moeller, Junbo Huang, Jan Reineke, Axel-Cyrille Ngonga~Ngomo, Muhammad Saleem, and Andreas Both. 2023.
\newblock \href {https://www.semantic-web-journal.net/content/qald-10-%E2%80%94-10th-challenge-question-answering-over-linked-data} {Qald-10 — the 10th challenge on question answering over linked data}.
\newblock \emph{Semantic Web – Interoperability, Usability, Applicability}.

\bibitem[{Vrande\v{c}i\'{c} and Kr\"{o}tzsch(2014)}]{wikidata}
Denny Vrande\v{c}i\'{c} and Markus Kr\"{o}tzsch. 2014.
\newblock \href {https://doi.org/10.1145/2629489} {Wikidata: a free collaborative knowledgebase}.
\newblock \emph{Commun. ACM}, 57(10):78–85.

\bibitem[{W3C(2006)}]{w3c:2006aa}
W3C. 2006.
\newblock {SPARQL} query language for {RDF -- Formal} definitions.
\newblock Accessible at http://www.w3.org/2001/sw/DataAccess/rq23/sparql-defns.html\#defn\_GroupGraphPattern.
\newblock Last accessed December 2015.

\bibitem[{Wylot et~al.(2018)Wylot, Hauswirth, Cudr{\'e}-Mauroux, and Sakr}]{Wylot:2018aa}
Marcin Wylot, Manfred Hauswirth, Philippe Cudr{\'e}-Mauroux, and Sherif Sakr. 2018.
\newblock \href {https://doi.org/10.1145/3177850} {Rdf data storage and query processing schemes: A survey}.
\newblock \emph{ACM Comput. Surv.}, 51(4):84:1--84:36.

\bibitem[{Xie et~al.(2022)Xie, Wu, Shi, Zhong, Scholak, Yasunaga, Wu, Zhong, Yin, Wang, Zhong, Wang, Li, Boyle, Ni, Yao, Radev, Xiong, Kong, Zhang, Smith, Zettlemoyer, and Yu}]{xie-etal-2022-unifiedskg}
Tianbao Xie, Chen~Henry Wu, Peng Shi, Ruiqi Zhong, Torsten Scholak, Michihiro Yasunaga, Chien-Sheng Wu, Ming Zhong, Pengcheng Yin, Sida~I. Wang, Victor Zhong, Bailin Wang, Chengzu Li, Connor Boyle, Ansong Ni, Ziyu Yao, Dragomir Radev, Caiming Xiong, Lingpeng Kong, and 4 others. 2022.
\newblock \href {https://doi.org/10.18653/v1/2022.emnlp-main.39} {{U}nified{SKG}: Unifying and multi-tasking structured knowledge grounding with text-to-text language models}.
\newblock In \emph{Proceedings of the 2022 Conference on Empirical Methods in Natural Language Processing}, pages 602--631, Abu Dhabi, United Arab Emirates. Association for Computational Linguistics.

\bibitem[{Yu et~al.(2023)Yu, Zhang, Ng, Zhu, Li, Wang, Hu, Wang, Wang, and Xiang}]{yu2023decaf}
Donghan Yu, Sheng Zhang, Patrick Ng, Henghui Zhu, Alexander~Hanbo Li, Jun Wang, Yiqun Hu, William~Yang Wang, Zhiguo Wang, and Bing Xiang. 2023.
\newblock \href {https://openreview.net/forum?id=XHc5zRPxqV9} {Dec{AF}: Joint decoding of answers and logical forms for question answering over knowledge bases}.
\newblock In \emph{The Eleventh International Conference on Learning Representations}.

\end{thebibliography}

\appendix



\end{document}